\documentclass[%
 reprint,
 superscriptaddress,
 amsmath,amssymb,
 aps,
 prl,
]{revtex4-2}

\usepackage{graphicx}%
\usepackage[colorlinks,breaklinks]{hyperref}%
\hypersetup{allcolors=blue}

\begin{document}

\title{Taming Landau level mixing in fractional quantum Hall states with deep learning}

\author{Yubing Qian}
\affiliation{School of Physics, Peking University, Beijing 100871, People’s Republic of China}%
\affiliation{ByteDance Research China, Fangheng Fashion Center, No. 27, North 3rd Ring West Road, Haidian District, Beijing 100098, People’s Republic of China}

\author{Tongzhou Zhao}
\email{tzzhao\_2022@iphy.ac.cn}
\thanks{Y.Q. and T.Z. contributed equally to this work.}
\affiliation{Institute of Physics, Chinese Academy of Sciences, Beijing 100190, China}

\author{Jianxiao Zhang}
\affiliation{Department of Physics, 104 Davey Lab, Pennsylvania State University, University Park, Pennsylvania 16802, USA}

\author{Tao Xiang}
\affiliation{Institute of Physics, Chinese Academy of Sciences, Beijing 100190, China}

\author{Xiang Li}
\email{lixiang.62770689@bytedance.com}
\affiliation{ByteDance Research China, Fangheng Fashion Center, No. 27, North 3rd Ring West Road, Haidian District, Beijing 100098, People’s Republic of China}

\author{Ji Chen}
\email{ji.chen@pku.edu.cn}
\affiliation{School of Physics, Peking University, Beijing 100871, People’s Republic of China}
\affiliation{Interdisciplinary Institute of Light-Element Quantum Materials, Frontiers Science Center for Nano-Optoelectronics, Peking University, Beijing 100871, People’s Republic of China}

\date{\today}

\begin{abstract}
Strong correlation brings a rich array of emergent phenomena, as well as a daunting challenge to theoretical physics study.
In condensed matter physics, the fractional quantum Hall effect is a prominent example of strong correlation, with Landau level mixing being one of the most challenging aspects to address using traditional computational methods.
Deep learning real-space neural network wavefunction methods have emerged as promising architectures to describe electron correlations in molecules and materials, but their power has not been fully tested for exotic quantum states. 
In this work, we employ real-space neural network wavefunction techniques to investigate fractional quantum Hall systems.
On both $1/3$ and $2/5$ filling systems, we achieve energies consistently lower than exact diagonalization results which only consider the lowest Landau level.
We also demonstrate that the real-space neural network wavefunction can naturally capture the extent of Landau level mixing up to a very high level, overcoming the limitations of traditional methods.
Our work underscores the potential of neural networks for future studies of strongly correlated systems and opens new avenues for exploring the rich physics of the fractional quantum Hall effect.
\end{abstract}
\maketitle

The fractional quantum Hall (FQH) effect is one of the most notable examples in condensed matter physics, highlighting the fascinating emergent phenomena driven by strong correlation effects and non-trivial topology~\cite{stormer_fractional_1999,laughlin_anomalous_1983}.
In FQH systems, the kinetic energy is quenched by the strong magnetic field, and the Coulomb interaction dominates the physics.
Consequently, the wavefunction of a FQH system can not be adiabatically connected to a simple, non-interacting state described by a single Slater determinant.
This complexity, combined with their intriguing topological properties, makes it challenging to represent these ground states efficiently using conventional methods.

Early studies usually assume infinitely strong magnetic fields, confining all electrons to the lowest Landau level (LLL).
However, in typical experiments, the Coulomb interaction strength is comparable to the cyclotron energy~\cite{sodemann_landau_2013}, and the Landau level mixing (LLM) can induce new physics.
Recent experiments have linked LLM to phenomena such as Wigner-crystal phase transitions~\cite{goldman_WC_1990, theibaut_fractional_2015},
spin transitions~\cite{eisenstein_evidence_1989},
and novel non-Abelian FQH states~\cite{luhman_observation_2008,wu_braiding_2014}.
Nevertheless, traditional numerical approaches have limitations in addressing LLM.
Exact diagonalization (ED) typically explicitly handles only the lowest Landau level (LLL) due to the large Hilbert space dimension~\cite{regnault_evidence_2017}, with very few studies extending it to two Landau levels~\cite{yoshioka_effect_1984}.
Density matrix renormalization group (DMRG) method
can include up to five Landau levels but struggles with states near the critical point with high entanglement~\cite{feiguin_density_2008,zaletel_infinite_2015}.
The fixed-phase diffusion Monte Carlo method (fp-DMC)
is limited by the accuracy of its phase approximation and does not provide an explicit wavefunction~\cite{ortiz_fp_dmc_1993,zhao_crystallization_2018,zhao_composite_2023}.

In recent years, deep learning methods have emerged as promising alternatives for studying strongly correlated systems~\cite{hermann_nnqmc-review_2023, qian_deep_2024}.
These methods have achieved remarkable success in representing the wavefunctions of various strongly correlated systems, including
lattice models~\cite{carleo_solving_2017,vicentini_variational_2019},
molecules~\cite{han_deepwf_2019, pfau_ferminet_2020, hermann_paulinet_2020, choo_fermionic_2020, glehn_psiformer_2023},
solids~\cite{yoshioka_rbm-solid_2021, li_electric_2024, li_deepsolid_2022},
electron gases~\cite{li_deepsolid_2022, wilson_wapnet_2023, cassella_discovering_2023, kim_ucf_2024},
and Moiré systems~\cite{li_emergent_2024, luo_simulating_2024}.
In these architectures, wavefunctions are presented in real space with neural networks, which can naturally include contributions from higher Landau levels within the ansatz.
This capability makes deep learning based on real-space neural networks a promising tool to break through the limitations of traditional methods and gain deeper insights into the ground states of FQH systems with LLM.

\begin{figure*}
    \centering
    \includegraphics{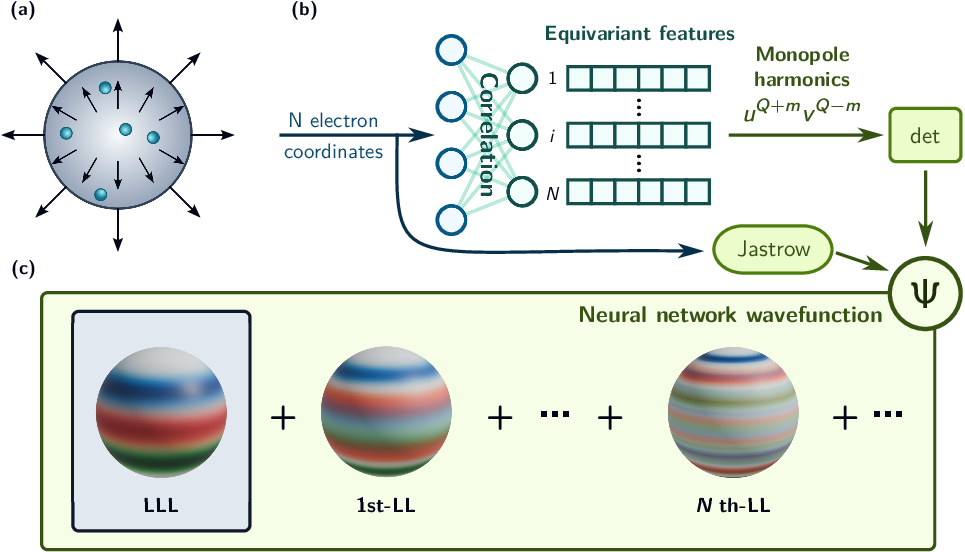}
    \caption{%
\textbf{(a)} Illustration of the system with spherical geometry. Electrons are confined to the spherical surface, with a magnetic monopole at the center.
\textbf{(b)} Deep learning computational framework of this work.
The neural network takes electron coordinates as inputs and outputs permutation-equivariant features that encode correlations of all electrons.
These features are then multiplied by monopole harmonics to form multi-electron orbitals.
The fermionic wavefunction is constructed by the determinant of the orbital matrix multiplied by a Jastrow factor.
\textbf{(c)} A schematic illustration of Landau level mixing.
Real-space neural network wavefunction goes beyond the lowest Landau level, and describes a mixture of an infinite number of Landau levels.
Only 3 orbitals are drawn per Landau level, colored as blue, red, and green.
}
    \label{fig:concept}
\end{figure*}

In this study, we employ real-space neural network methods to investigate FQH systems in spherical geometry~\cite{haldane_fractional_1983}.
In the literature, disk~\cite{laughlin_anomalous_1983} and torus~\cite{yoshioka_ground_1983} geometries are also employed to study FQH effects, compared to which the spherical geometry has several advantages, including the edgeless structure, the well-defined filling factor and the simplicity in mathematical treatment.
As shown in Fig.~\ref{fig:concept}a, $N$ spin-polarized electrons are confined on a spherical surface, with a magnetic monopole placed at the center.
The monopole creates a total flux of $2Q\phi_0$ through the surface, where $\phi_0$ is the flux quantum and $2Q$ is an integer.
The radius of the sphere $R$ is given by $\sqrt{Q}\ell$, where $\ell=\sqrt{\hbar c/eB}$ is the magnetic length, and $B$ is the strength of the uniform magnetic field on the sphere.
The flux-particle relationship on the sphere is determined by $2Q=N/\nu -\mathcal{S}$, where $\nu$ is the filling factor and $\mathcal{S}$ is the shift, which characterizes the topological properties of the corresponding FQH state~\cite{wen_shift_1992}.

The Hamiltonian on the spherical geometry is formulated as:
\begin{equation}
    \hat H = \sum_i\frac{\ell^2 \hbar \omega_c}{2R^2}|\hat{\boldsymbol{\Lambda}}_i|^2 + \frac{e^2}{\epsilon} \sum_{i < j} \frac{1}{|\mathbf{r}_i - \mathbf{r}_j|},
\end{equation}
where $\omega_c=eB/mc$ is the cyclotron frequency,
$m$ is the band mass of the electrons,
$\epsilon$ is the dielectric constant of the material,
and $\mathbf{r}_i$ is the coordinate of the $i$-th electron.
$\hat{\boldsymbol{\Lambda}}_i$ is proportional to the canonical momentum tangential to the surface:
\begin{equation}
    |\hat{\boldsymbol{\Lambda}}_i|^2=-\frac{1}{\sin \theta_i}\frac{\partial}{\partial \theta_i}\sin\theta_i\frac{\partial}{\partial \theta_i}
    +\left(Q\cot\theta_i+\frac{\mathrm{i}}{\sin \theta_i}\frac{\partial}{\partial \phi_i}\right)^2.
\end{equation}
We define
$\kappa=(e^2/\epsilon \ell) / (\hbar \omega_c)$ as a measure of the interaction strength, which also effectively tunes the LLM.
All energies shown are in unit of $\hbar \omega_c \kappa$.

\begin{figure*}
    \centering
    \includegraphics{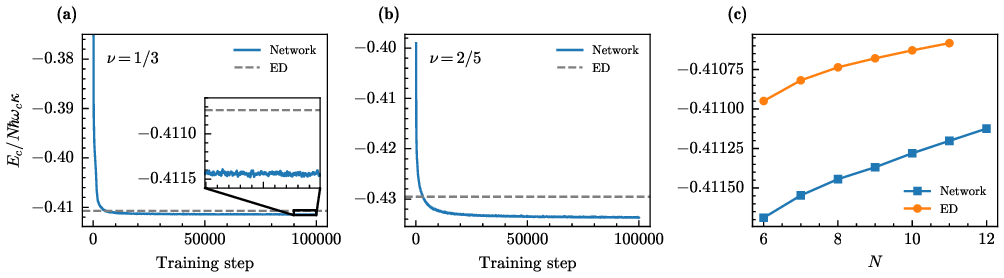}
    \caption{%
\textbf{(a)} The training curve of a $\nu = 1/3$ system with $N=8$, $2Q=21$, and $\kappa=1$.
A moving window average of 100 steps is applied and outliers are removed.
The dashed line is the energy result from ED.
\textbf{(b)} The training curve of a $\nu=2/5$ system, where $N=8$, $2Q=16$ and $\kappa=1$.
\textbf{(c)} ED (orange) and neural network (blue) energy results with different number of electrons for $\nu = 1/3$ filling.
The energy error bars are smaller than the size of the markers.
The energy ($E_c$) includes the contribution from background charge~\cite{jain_composite_2007}.
In addition, it is shifted by $N\omega_c/2$ and a density correction to reduce the system size dependence is applied~\cite{morf_microscopic_1986, jain_composite_2007}.
The energy unit is $\hbar \omega_c \kappa$.
    }
    \label{fig:train-and-energy}
\end{figure*}

Crafting a valid and expressive neural network wavefunction ansatz is essential for obtaining fractional quantum Hall states with deep learning.
To achieve this, we design a trial wavefunction $\psi_T^{\text{nn}}$ as follows:
\begin{equation}
    \psi_T^\text{nn}(\mathbf{r}) = \mathrm{e}^{\mathcal{J}(\mathbf{r})} \operatorname{det} [\varphi_{i}(\mathbf{r}_j;\{\mathbf{r}_{\ne j}\})],
\end{equation}
where $\mathbf{r}$ denotes the coordinates of all electrons, and the notation $\{\mathbf{r}_{\ne j}\}$ indicates that the permutation of other electrons does not change the output (Fig.~\ref{fig:concept}b).
The Jastrow factor
\begin{equation}
    \mathcal{J}(\mathbf{r}) = -\frac{1}{4}\sum_{i<j} \frac{a^2}{a+|\mathbf{r}_i - \mathbf{r}_j|}
\end{equation}
is included to satisfy the electron--electron Coulomb cusp condition~\cite{kato_cusp_1957}, where $a$ is a free parameter.
And the multi-electron orbitals $\varphi_{i}(\mathbf{r}_j;\{\mathbf{r}_{\ne j}\})$ on the sphere are given by
\begin{equation}\label{eq:orbital}
    \varphi_{i}(\mathbf{r}_j;\{\mathbf{r}_{\ne j}\}) = \sum_{k,m} w_{ikm} f_{k}(\mathbf{r}_j;\{\mathbf{r}_{\ne j}\}) u_j^{Q+m} v_j^{Q-m},
\end{equation}
where $w_{ikm}$ are complex parameters and $u_j^{Q+m} v_j^{Q-m}$ are monopole harmonics~\cite{wu_dirac_1976} on the LLL, with $m=-|Q|,-|Q|+1,\dots,|Q|$.
The spinor coordinates of the $j$-th electron are given by
$u_j=\cos(\theta_j/2) \mathrm{e}^{\mathrm{i} \phi_j/2}$ and $v_j=\sin(\theta_j/2) \mathrm{e}^{-\mathrm{i} \phi_j/2}$.
The use of monopole harmonics largely avoids the divergence problem in local energy at the poles of the sphere originating from Dirac strings, and the physical information such as the correlations and topological properties is encoded in the neural network $f_{k}(\mathbf{r}_j;\{\mathbf{r}_{\ne j}\})$.
Here, the Psiformer neural network architecture~\cite{glehn_psiformer_2023} is adopted, which is capable of modeling strongly correlated electrons.

With the ansatz constructed, we then employ the variational Monte Carlo process to optimize the wavefunction.
The loss function during training is chosen as the energy $E_v$ evaluated using Monte Carlo integration.
The Kronecker-Factored Approximate Curvature method~\cite{martens_kfac_2015} is used to optimize the parameters.
With the expressiveness of the neural network, it is possible to describe the quantum Hall states using a single determinant constructed with multi-electron orbitals $\varphi_{i}(\mathbf{r}_j;\{\mathbf{r}_{\ne j}\})$.
Moreover, the neural network wavefunction can naturally include the effects of LLM (Fig. \ref{fig:concept}c), surpassing the calculations that only include few Landau levels.

We first apply the neural network based variational Monte Carlo (NNVMC) to investigate the most robust FQH effect at $\nu=1/3$, for which the Laughlin wavefunction provides a good description of the ground state~\cite{laughlin_anomalous_1983}.
A simulation containing $N=8$ electrons with flux $2Q=21$ and $\kappa=1$ is shown in Fig.~\ref{fig:train-and-energy}a.
The energy obtained from the neural network reaches lower than the ED result, which only considers the LLL.
This reflects the contribution of higher Landau levels due to the Coulomb interaction.
We also use our approach to study the more complex $\nu=2/5$ system, for which the composite fermion wavefunction of Jain was known to be a good ansatz~\cite{jain_composite-fermion_1989}.
As demonstrated in Fig.~\ref{fig:train-and-energy}b, although the neural network converges more slowly compared to the $\nu=1/3$ state, it still reaches a much lower energy than the ED result.
This indicates that the real-space neural network captures electron correlations consistently and can be successfully applied to study FQH systems of different fillings.

To further validate the robustness of our approach, we examine the energy as a function of the number of electrons for the $\nu=1/3$ state.
As shown in Fig.~\ref{fig:train-and-energy}c, the neural network results from 6 to 12 electrons are consistently lower than the ED results.
This demonstrates the neural network's ability to accurately capture the ground state energies across different system sizes.
Moreover, it is worth noting that the system with $N=12$ electrons is inaccessible to ED due to the exponentially growing Hilbert space.
However, NNVMC can handle this larger system size and the results align well with the trend observed for smaller numbers of electrons.
This further underscores the scalability and effectiveness of neural network methods in exploring larger FQH systems.

\begin{figure*}
    \centering
    \includegraphics{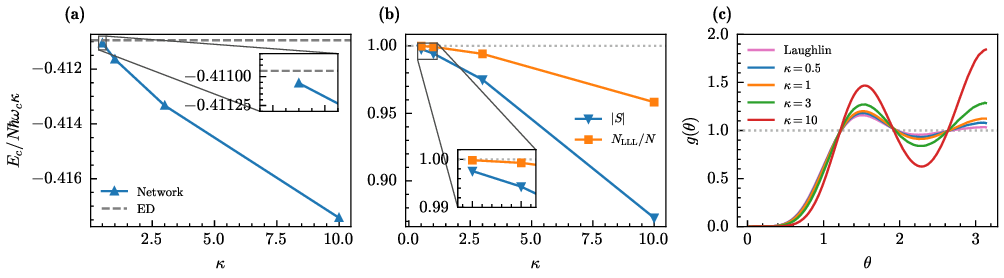}
    \caption{%
\textbf{(a)} Energy per electron as a function of Landau level mixing parameter $\kappa$.
ED is performed on the lowest Landau level and is the same under different $\kappa$, so it is shown as a horizontal dashed line.
\textbf{(b)} Overlap modulus of the neural network wavefunction with the Laughlin wavefunction (blue) and the ratio of electrons on the lowest Landau level (orange) as a function of $\kappa$.
The error bars are smaller than the size of the markers.
\textbf{(c)} Pair correlation function $g(\theta)$ for different values of $\kappa$.
All simulations are performed with $N=6$ electrons, flux $2Q=15$.
    }
    \label{fig:ground-llm}
\end{figure*}

The above results are obtained with $\kappa=1$, where LLM is mild.
We now investigate the $\nu=1/3$ system with $N=6$ electrons and varying $\kappa$ to examine whether our neural network architecture is able to learn FQH states for different levels of LLM.
As illustrated in Fig.~\ref{fig:ground-llm}a, for a weak interaction strength, $\kappa=0.5$, where the lowest Landau level is dominant, the energy obtained from NNVMC is slightly lower than the ED result.
This shows that the neural network can reliably capture the contribution of higher Landau levels even when LLM is small.
As $\kappa$ increases, LLM becomes more pronounced and the contributions from higher Landau levels become more significant.
Consequently, ED calculations considering only the LLL would significantly overestimate the energy, hence NNVMC can reach much lower energy than ED.
As expected, the gap between NNVMC and ED increases monotonously as a function of LLM.

\begin{figure*}
    \centering
    \includegraphics{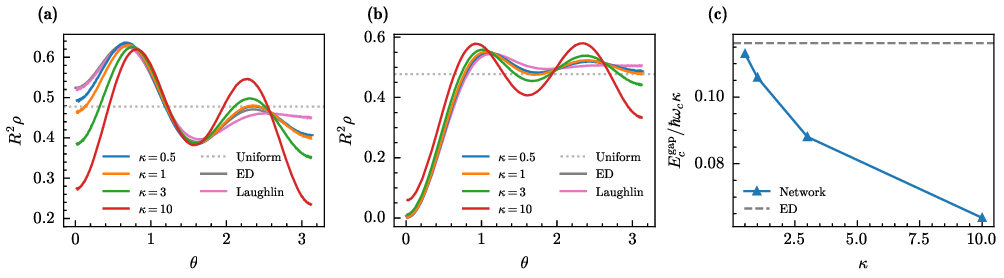}
    \caption{%
\textbf{(a)} Density of quasiparticle states derived from ED, the Laughlin wavefunction, and neural network wavefunctions with various $\kappa$.
The quasiparticle is localized at the north pole.
The density is scaled by the square of the radius $R$ of the spherical geometry.
\textbf{(b)} Similar to (a), but for the density of quasihole states.
\textbf{(c)} Corrected transport gap as a function of the Landau level mixing parameter $\kappa$.
The error bars are smaller than the size of the markers.
The calculations are performed for $N=6$ electrons, with flux $2Q=14,15,16$ corresponding to quasiparticle, ground state, and quasihole states, respectively.
    }
    \label{fig:excite-llm}
\end{figure*}

To further reveal the LLM effects, we monitor the overlap $S$ between the neural network wavefunction and the Laughlin wavefunction, which measures the similarity between them.
The overlap is defined as:
\begin{equation}\label{eq:overlap}
    S=\frac{\langle \psi^\text{nn} | \psi^\text{Laughlin}\rangle}
    {\left(\langle\psi^\text{nn} |\psi^\text{nn}\rangle\langle \psi^\text{Laughlin}|\psi^\text{Laughlin}\rangle\right)^{1/2}}.
\end{equation}
In Fig.~\ref{fig:ground-llm}b, we plot the modulus of the overlap ($|S|$).
When $\kappa$ is small, a large $|S|$ is obtained,
since the Laughlin wavefunction is an excellent approximation of the ground state wavefunction in the absence of LLM.
As LLM grows stronger, the modulus of overlap $|S|$ decreases, reasonably reflecting the influence of LLM.

The above results do not directly indicate whether the neural network wavefunction includes contributions from higher Landau levels.
To gain deeper insights, we also examine the number of electrons on the LLL, denoted as $N_\text{LLL}$.
Our results in Fig.~\ref{fig:ground-llm}b show that $N_\text{LLL}$ follows the same trend as $|S|$, being large at small $\kappa$ and decreasing with larger $\kappa$.
Although approximately 96\% of the electrons remain on the LLL even at $\kappa=10$, the contribution from the remaining 4\% of electrons in higher Landau levels is not negligible, as evidenced by the energy results.
This highlights the significant role of higher Landau levels in the presence of LLM.

The neural network can also model the behavior of the pair correlation function (PCF) $g(\theta)$, which describes the probability of finding two electrons with an angular separation of $\theta$ and reveals the correlations between electrons.
The $\nu=1/3$ FQH state is known to resemble a liquid~\cite{kamilla_fermi-sea-like_1997}, as 
illustrated by the PCF of the Laughlin wavefunction in Fig.~\ref{fig:ground-llm}c.
The PCF from the neural network wavefunction closely matches that of the Laughlin wavefunction when $\kappa=0.5$.
As LLM increases, the PCF becomes more structured.
Eventually, at $\kappa=10$, the PCF no longer exhibits a decaying trend and displays a prominent peak at $\theta=\pi$, indicating a phase transition away from the liquid phase~\cite{yannouleas_two-dimensional_2003,Yannouleas_trial_2002,shibata_ground-state_dmrg_2001}.
However, the current system size in our approach is not sufficient to definitively establish the phase boundary.
This is also evident in the subsequent energy gap calculations, where the charge gap persists up to $\kappa=10$, as shown in Fig.~\ref{fig:excite-llm}c.
Nevertheless, the transition trend observed in our calculations aligns with the classical limit $\kappa\to\infty$, where the system is known to form a classical Wigner crystal with electrons arranged in a triangular lattice~\cite{thomson_structure_1904,e_wigner_interaction_1934}.

Besides the ground state, the corresponding quasiparticle and quasihole excitations in FQH systems also exhibit interesting physics.
These excitations carry fractional electric charges, a remarkable departure from the usual integer charges, and they exhibit anyonic braiding statistics, which are neither bosonic nor fermionic.
On the spherical geometry, the quasiparticle and quasihole states are actually \textit{ground states} for different flux values $2Q^\text{qp/qh} = 3(N-1)\mp1$.
To reveal the charge density for these excitations, we need to break the degeneracy of different excitations by adding a penalty term $\beta (\hat L_z - m_t)^2$ to the loss function during the network training, where $L_z$ is the total angular momentum in the $z$-direction, $m_t$ is the target $L_z$, and $\beta$ is a tunable hyperparameter.
The training is first converged with $\beta=0$, and then we turn on $\beta$ to select the state whose $L_z = m_t$.
In Fig.~\ref{fig:excite-llm}a--b, we plot the charge density for quasiparticle and quasihole excitations from NNVMC for the $\nu=1/3$ system with $\kappa=1$, scaled by the square of sphere radius $R$.
Notably, since the Laughlin wavefunction captures short-range physics~\cite{haldane_fractional_1983}, the excitations are more localized compared to the ED and neural network results.
For small $\kappa$, the neural network results agree well with the ED results, further demonstrating that our neural network wavefunction accurately captures the physics of the FQH system.
As $\kappa$ increases, the density fluctuations are enhanced compared to the ED results, consistent with the behavior of the PCF shown in Fig.~\ref{fig:ground-llm}c.
At a large $\kappa=10$, the quasihole excitation density near the center is no longer close to zero, indicating a phase transition at high $\kappa$.

Having demonstrated the accuracy of neural network wavefunction, we can now study the transport gap of FQH systems and the effects of LLM on it, which are important topics that pose challenges to the theoretical community~\cite{haldane_finite-size_1985,morf_excitation_2002,zhao_revisiting_2022}.
The transport gap corresponds to the energy cost of moving a quasiparticle far from the system, leaving a quasihole behind.
This gap can be measured in finite-temperature transport experiments~\cite{boebinger_magnetic_1985,du_experimental_1993}.
It also reflects the composite fermion mass and can be deduced by the composite fermion Chern--Simons field theory~\cite{halperin_theory_1993,murthy_cs-review_2003,yu_effective_1998,Praz_effective_2007}. 
The transport gap is defined as $E^\text{gap}=E_v^\text{qp} + E_v^\text{qh} - 2 E_v$, where $E^\text{qp}$ and $E^\text{qh}$ are quasiparticle and quasihole energies, respectively.
Similar to Fig.~\ref{fig:train-and-energy}, we denote the corrected gap as $E^\text{gap}_c$.
As shown in Fig.~\ref{fig:excite-llm}c, for small values of $\kappa$, the transport gap aligns closely with the value obtained from ED.
As $\kappa$ increases, the transport gap decreases progressively.
However, even at $\kappa=10$, the transport gap does not fully close, indicating that the system remains in a gapped phase despite the significant influence of LLM.
These results are all consistent with previous studies~\cite{,melik-alaverdian_composite_1995, murthy_hamiltonian_2002, zhao_revisiting_2022}.

In conclusion, we have demonstrated the effectiveness of deep learning methods in investigating FQH systems, capturing contributions from higher Landau levels. 
We employ the spherical geometry and examine the bulk properties and quasiparticle excitations of the $\nu=1/3$ and $\nu=2/5$ states, where we show that real-space neural networks can learn more accurate wavefunctions than those obtained by exact diagonalization based on the lowest Landau levels. 
In a concurrent study, a similar approach to this work is devised to study FQH on disk geometry, where similar conclusions were drawn, further highlighting the strengths of neural network methods~\cite{teng_solving_2024}.
Together with this work, these successes highlight the potential of neural networks for future explorations, including investigations into additional filling factors and other topological properties.
Our results also show hints of the phase transition, but further investigations are desired to elucidate the issue.
In the future, deep learning frameworks can be further designed to encode key physical information, which may lead to more efficient solutions and more advancements in addressing challenging questions resulting from strong correlations.

\section*{Acknowledgments}
The authors would like to thank Nicholas Regnault, Jiequn Han, and Xi Dai for their helpful discussions, and ByteDance Research Group for inspiration and encouragement.
J.C. acknowledges supports from the National Key R\&D Program of China under Grant No. 2021YFA1400500, the National Natural Science Foundation of China under Grant No. 12334003, and the Beijing Municipal Natural Science Foundation under Grant No. JQ22001.
T.X. acknowledges supports by the NSFC grant No. 12488201.
T.Z. acknowledges supports by the China Postdoctoral Science Foundation grant No. 2023M743742.

\bibliography{ref}

\begin{thebibliography}{61}%
\makeatletter
\providecommand \@ifxundefined [1]{%
 \@ifx{#1\undefined}
}%
\providecommand \@ifnum [1]{%
 \ifnum #1\expandafter \@firstoftwo
 \else \expandafter \@secondoftwo
 \fi
}%
\providecommand \@ifx [1]{%
 \ifx #1\expandafter \@firstoftwo
 \else \expandafter \@secondoftwo
 \fi
}%
\providecommand \natexlab [1]{#1}%
\providecommand \enquote  [1]{``#1''}%
\providecommand \bibnamefont  [1]{#1}%
\providecommand \bibfnamefont [1]{#1}%
\providecommand \citenamefont [1]{#1}%
\providecommand \href@noop [0]{\@secondoftwo}%
\providecommand \href [0]{\begingroup \@sanitize@url \@href}%
\providecommand \@href[1]{\@@startlink{#1}\@@href}%
\providecommand \@@href[1]{\endgroup#1\@@endlink}%
\providecommand \@sanitize@url [0]{\catcode `\\12\catcode `\$12\catcode `\&12\catcode `\#12\catcode `\^12\catcode `\_12\catcode `\%12\relax}%
\providecommand \@@startlink[1]{}%
\providecommand \@@endlink[0]{}%
\providecommand \url  [0]{\begingroup\@sanitize@url \@url }%
\providecommand \@url [1]{\endgroup\@href {#1}{\urlprefix }}%
\providecommand \urlprefix  [0]{URL }%
\providecommand \Eprint [0]{\href }%
\providecommand \doibase [0]{https://doi.org/}%
\providecommand \selectlanguage [0]{\@gobble}%
\providecommand \bibinfo  [0]{\@secondoftwo}%
\providecommand \bibfield  [0]{\@secondoftwo}%
\providecommand \translation [1]{[#1]}%
\providecommand \BibitemOpen [0]{}%
\providecommand \bibitemStop [0]{}%
\providecommand \bibitemNoStop [0]{.\EOS\space}%
\providecommand \EOS [0]{\spacefactor3000\relax}%
\providecommand \BibitemShut  [1]{\csname bibitem#1\endcsname}%
\let\auto@bib@innerbib\@empty
\bibitem [{\citenamefont {Stormer}\ \emph {et~al.}(1999)\citenamefont {Stormer}, \citenamefont {Tsui},\ and\ \citenamefont {Gossard}}]{stormer_fractional_1999}%
  \BibitemOpen
  \bibfield  {author} {\bibinfo {author} {\bibfnamefont {H.~L.}\ \bibnamefont {Stormer}}, \bibinfo {author} {\bibfnamefont {D.~C.}\ \bibnamefont {Tsui}},\ and\ \bibinfo {author} {\bibfnamefont {A.~C.}\ \bibnamefont {Gossard}},\ }\href {https://doi.org/10.1103/RevModPhys.71.S298} {\bibfield  {journal} {\bibinfo  {journal} {Reviews of Modern Physics}\ }\textbf {\bibinfo {volume} {71}},\ \bibinfo {pages} {S298} (\bibinfo {year} {1999})}\BibitemShut {NoStop}%
\bibitem [{\citenamefont {Laughlin}(1983)}]{laughlin_anomalous_1983}%
  \BibitemOpen
  \bibfield  {author} {\bibinfo {author} {\bibfnamefont {R.~B.}\ \bibnamefont {Laughlin}},\ }\href {https://doi.org/10.1103/PhysRevLett.50.1395} {\bibfield  {journal} {\bibinfo  {journal} {Phys. Rev. Lett.}\ }\textbf {\bibinfo {volume} {50}},\ \bibinfo {pages} {1395} (\bibinfo {year} {1983})}\BibitemShut {NoStop}%
\bibitem [{\citenamefont {Sodemann}\ and\ \citenamefont {MacDonald}(2013)}]{sodemann_landau_2013}%
  \BibitemOpen
  \bibfield  {author} {\bibinfo {author} {\bibfnamefont {I.}~\bibnamefont {Sodemann}}\ and\ \bibinfo {author} {\bibfnamefont {A.~H.}\ \bibnamefont {MacDonald}},\ }\href {https://doi.org/10.1103/PhysRevB.87.245425} {\bibfield  {journal} {\bibinfo  {journal} {Physical Review B}\ }\textbf {\bibinfo {volume} {87}},\ \bibinfo {pages} {245425} (\bibinfo {year} {2013})}\BibitemShut {NoStop}%
\bibitem [{\citenamefont {Goldman}\ \emph {et~al.}(1990)\citenamefont {Goldman}, \citenamefont {Santos}, \citenamefont {Shayegan},\ and\ \citenamefont {Cunningham}}]{goldman_WC_1990}%
  \BibitemOpen
  \bibfield  {author} {\bibinfo {author} {\bibfnamefont {V.~J.}\ \bibnamefont {Goldman}}, \bibinfo {author} {\bibfnamefont {M.}~\bibnamefont {Santos}}, \bibinfo {author} {\bibfnamefont {M.}~\bibnamefont {Shayegan}},\ and\ \bibinfo {author} {\bibfnamefont {J.~E.}\ \bibnamefont {Cunningham}},\ }\href {https://doi.org/10.1103/PhysRevLett.65.2189} {\bibfield  {journal} {\bibinfo  {journal} {Phys. Rev. Lett.}\ }\textbf {\bibinfo {volume} {65}},\ \bibinfo {pages} {2189} (\bibinfo {year} {1990})}\BibitemShut {NoStop}%
\bibitem [{\citenamefont {Thiebaut}\ \emph {et~al.}(2015)\citenamefont {Thiebaut}, \citenamefont {Regnault},\ and\ \citenamefont {Goerbig}}]{theibaut_fractional_2015}%
  \BibitemOpen
  \bibfield  {author} {\bibinfo {author} {\bibfnamefont {N.}~\bibnamefont {Thiebaut}}, \bibinfo {author} {\bibfnamefont {N.}~\bibnamefont {Regnault}},\ and\ \bibinfo {author} {\bibfnamefont {M.~O.}\ \bibnamefont {Goerbig}},\ }\href {https://doi.org/10.1103/PhysRevB.92.245401} {\bibfield  {journal} {\bibinfo  {journal} {Phys. Rev. B}\ }\textbf {\bibinfo {volume} {92}},\ \bibinfo {pages} {245401} (\bibinfo {year} {2015})}\BibitemShut {NoStop}%
\bibitem [{\citenamefont {Eisenstein}\ \emph {et~al.}(1989)\citenamefont {Eisenstein}, \citenamefont {Stormer}, \citenamefont {Pfeiffer},\ and\ \citenamefont {West}}]{eisenstein_evidence_1989}%
  \BibitemOpen
  \bibfield  {author} {\bibinfo {author} {\bibfnamefont {J.~P.}\ \bibnamefont {Eisenstein}}, \bibinfo {author} {\bibfnamefont {H.~L.}\ \bibnamefont {Stormer}}, \bibinfo {author} {\bibfnamefont {L.}~\bibnamefont {Pfeiffer}},\ and\ \bibinfo {author} {\bibfnamefont {K.~W.}\ \bibnamefont {West}},\ }\href {https://doi.org/10.1103/PhysRevLett.62.1540} {\bibfield  {journal} {\bibinfo  {journal} {Phys. Rev. Lett.}\ }\textbf {\bibinfo {volume} {62}},\ \bibinfo {pages} {1540} (\bibinfo {year} {1989})}\BibitemShut {NoStop}%
\bibitem [{\citenamefont {Luhman}\ \emph {et~al.}(2008)\citenamefont {Luhman}, \citenamefont {Pan}, \citenamefont {Tsui}, \citenamefont {Pfeiffer}, \citenamefont {Baldwin},\ and\ \citenamefont {West}}]{luhman_observation_2008}%
  \BibitemOpen
  \bibfield  {author} {\bibinfo {author} {\bibfnamefont {D.~R.}\ \bibnamefont {Luhman}}, \bibinfo {author} {\bibfnamefont {W.}~\bibnamefont {Pan}}, \bibinfo {author} {\bibfnamefont {D.~C.}\ \bibnamefont {Tsui}}, \bibinfo {author} {\bibfnamefont {L.~N.}\ \bibnamefont {Pfeiffer}}, \bibinfo {author} {\bibfnamefont {K.~W.}\ \bibnamefont {Baldwin}},\ and\ \bibinfo {author} {\bibfnamefont {K.~W.}\ \bibnamefont {West}},\ }\href {https://doi.org/10.1103/PhysRevLett.101.266804} {\bibfield  {journal} {\bibinfo  {journal} {Phys. Rev. Lett.}\ }\textbf {\bibinfo {volume} {101}},\ \bibinfo {pages} {266804} (\bibinfo {year} {2008})}\BibitemShut {NoStop}%
\bibitem [{\citenamefont {Wu}\ \emph {et~al.}(2014)\citenamefont {Wu}, \citenamefont {Estienne}, \citenamefont {Regnault},\ and\ \citenamefont {Bernevig}}]{wu_braiding_2014}%
  \BibitemOpen
  \bibfield  {author} {\bibinfo {author} {\bibfnamefont {Y.-L.}\ \bibnamefont {Wu}}, \bibinfo {author} {\bibfnamefont {B.}~\bibnamefont {Estienne}}, \bibinfo {author} {\bibfnamefont {N.}~\bibnamefont {Regnault}},\ and\ \bibinfo {author} {\bibfnamefont {B.~A.}\ \bibnamefont {Bernevig}},\ }\href {https://doi.org/10.1103/PhysRevLett.113.116801} {\bibfield  {journal} {\bibinfo  {journal} {Phys. Rev. Lett.}\ }\textbf {\bibinfo {volume} {113}},\ \bibinfo {pages} {116801} (\bibinfo {year} {2014})}\BibitemShut {NoStop}%
\bibitem [{\citenamefont {Regnault}\ \emph {et~al.}(2017)\citenamefont {Regnault}, \citenamefont {Maciejko}, \citenamefont {Kivelson},\ and\ \citenamefont {Sondhi}}]{regnault_evidence_2017}%
  \BibitemOpen
  \bibfield  {author} {\bibinfo {author} {\bibfnamefont {N.}~\bibnamefont {Regnault}}, \bibinfo {author} {\bibfnamefont {J.}~\bibnamefont {Maciejko}}, \bibinfo {author} {\bibfnamefont {S.~A.}\ \bibnamefont {Kivelson}},\ and\ \bibinfo {author} {\bibfnamefont {S.~L.}\ \bibnamefont {Sondhi}},\ }\href {https://doi.org/10.1103/PhysRevB.96.035150} {\bibfield  {journal} {\bibinfo  {journal} {Phys. Rev. B}\ }\textbf {\bibinfo {volume} {96}},\ \bibinfo {pages} {035150} (\bibinfo {year} {2017})}\BibitemShut {NoStop}%
\bibitem [{\citenamefont {Yoshioka}(1984)}]{yoshioka_effect_1984}%
  \BibitemOpen
  \bibfield  {author} {\bibinfo {author} {\bibfnamefont {D.}~\bibnamefont {Yoshioka}},\ }\href {https://doi.org/10.1143/JPSJ.53.3740} {\bibfield  {journal} {\bibinfo  {journal} {Journal of the Physical Society of Japan}\ }\textbf {\bibinfo {volume} {53}},\ \bibinfo {pages} {3740} (\bibinfo {year} {1984})}\BibitemShut {NoStop}%
\bibitem [{\citenamefont {Feiguin}\ \emph {et~al.}(2008)\citenamefont {Feiguin}, \citenamefont {Rezayi}, \citenamefont {Nayak},\ and\ \citenamefont {Das~Sarma}}]{feiguin_density_2008}%
  \BibitemOpen
  \bibfield  {author} {\bibinfo {author} {\bibfnamefont {A.~E.}\ \bibnamefont {Feiguin}}, \bibinfo {author} {\bibfnamefont {E.}~\bibnamefont {Rezayi}}, \bibinfo {author} {\bibfnamefont {C.}~\bibnamefont {Nayak}},\ and\ \bibinfo {author} {\bibfnamefont {S.}~\bibnamefont {Das~Sarma}},\ }\href {https://doi.org/10.1103/PhysRevLett.100.166803} {\bibfield  {journal} {\bibinfo  {journal} {Physical Review Letters}\ }\textbf {\bibinfo {volume} {100}},\ \bibinfo {pages} {166803} (\bibinfo {year} {2008})}\BibitemShut {NoStop}%
\bibitem [{\citenamefont {Zaletel}\ \emph {et~al.}(2015)\citenamefont {Zaletel}, \citenamefont {Mong}, \citenamefont {Pollmann},\ and\ \citenamefont {Rezayi}}]{zaletel_infinite_2015}%
  \BibitemOpen
  \bibfield  {author} {\bibinfo {author} {\bibfnamefont {M.~P.}\ \bibnamefont {Zaletel}}, \bibinfo {author} {\bibfnamefont {R.~S.~K.}\ \bibnamefont {Mong}}, \bibinfo {author} {\bibfnamefont {F.}~\bibnamefont {Pollmann}},\ and\ \bibinfo {author} {\bibfnamefont {E.~H.}\ \bibnamefont {Rezayi}},\ }\href {https://doi.org/10.1103/PhysRevB.91.045115} {\bibfield  {journal} {\bibinfo  {journal} {Physical Review B}\ }\textbf {\bibinfo {volume} {91}},\ \bibinfo {pages} {045115} (\bibinfo {year} {2015})}\BibitemShut {NoStop}%
\bibitem [{\citenamefont {Ortiz}\ \emph {et~al.}(1993)\citenamefont {Ortiz}, \citenamefont {Ceperley},\ and\ \citenamefont {Martin}}]{ortiz_fp_dmc_1993}%
  \BibitemOpen
  \bibfield  {author} {\bibinfo {author} {\bibfnamefont {G.}~\bibnamefont {Ortiz}}, \bibinfo {author} {\bibfnamefont {D.~M.}\ \bibnamefont {Ceperley}},\ and\ \bibinfo {author} {\bibfnamefont {R.~M.}\ \bibnamefont {Martin}},\ }\href {https://doi.org/10.1103/PhysRevLett.71.2777} {\bibfield  {journal} {\bibinfo  {journal} {Physical Review Letters}\ }\textbf {\bibinfo {volume} {71}},\ \bibinfo {pages} {2777} (\bibinfo {year} {1993})}\BibitemShut {NoStop}%
\bibitem [{\citenamefont {Zhao}\ \emph {et~al.}(2018)\citenamefont {Zhao}, \citenamefont {Zhang},\ and\ \citenamefont {Jain}}]{zhao_crystallization_2018}%
  \BibitemOpen
  \bibfield  {author} {\bibinfo {author} {\bibfnamefont {J.}~\bibnamefont {Zhao}}, \bibinfo {author} {\bibfnamefont {Y.}~\bibnamefont {Zhang}},\ and\ \bibinfo {author} {\bibfnamefont {J.~K.}\ \bibnamefont {Jain}},\ }\href {https://doi.org/10.1103/PhysRevLett.121.116802} {\bibfield  {journal} {\bibinfo  {journal} {Phys. Rev. Lett.}\ }\textbf {\bibinfo {volume} {121}},\ \bibinfo {pages} {116802} (\bibinfo {year} {2018})}\BibitemShut {NoStop}%
\bibitem [{\citenamefont {Zhao}\ \emph {et~al.}(2023)\citenamefont {Zhao}, \citenamefont {Balram},\ and\ \citenamefont {Jain}}]{zhao_composite_2023}%
  \BibitemOpen
  \bibfield  {author} {\bibinfo {author} {\bibfnamefont {T.}~\bibnamefont {Zhao}}, \bibinfo {author} {\bibfnamefont {A.~C.}\ \bibnamefont {Balram}},\ and\ \bibinfo {author} {\bibfnamefont {J.~K.}\ \bibnamefont {Jain}},\ }\href {https://doi.org/10.1103/PhysRevLett.130.186302} {\bibfield  {journal} {\bibinfo  {journal} {Phys. Rev. Lett.}\ }\textbf {\bibinfo {volume} {130}},\ \bibinfo {pages} {186302} (\bibinfo {year} {2023})}\BibitemShut {NoStop}%
\bibitem [{\citenamefont {Hermann}\ \emph {et~al.}(2023)\citenamefont {Hermann}, \citenamefont {Spencer}, \citenamefont {Choo}, \citenamefont {Mezzacapo}, \citenamefont {Foulkes}, \citenamefont {Pfau}, \citenamefont {Carleo},\ and\ \citenamefont {No{\'e}}}]{hermann_nnqmc-review_2023}%
  \BibitemOpen
  \bibfield  {author} {\bibinfo {author} {\bibfnamefont {J.}~\bibnamefont {Hermann}}, \bibinfo {author} {\bibfnamefont {J.}~\bibnamefont {Spencer}}, \bibinfo {author} {\bibfnamefont {K.}~\bibnamefont {Choo}}, \bibinfo {author} {\bibfnamefont {A.}~\bibnamefont {Mezzacapo}}, \bibinfo {author} {\bibfnamefont {W.~M.~C.}\ \bibnamefont {Foulkes}}, \bibinfo {author} {\bibfnamefont {D.}~\bibnamefont {Pfau}}, \bibinfo {author} {\bibfnamefont {G.}~\bibnamefont {Carleo}},\ and\ \bibinfo {author} {\bibfnamefont {F.}~\bibnamefont {No{\'e}}},\ }\href {https://doi.org/10.1038/s41570-023-00516-8} {\bibfield  {journal} {\bibinfo  {journal} {Nature Reviews Chemistry}\ }\textbf {\bibinfo {volume} {7}},\ \bibinfo {pages} {692} (\bibinfo {year} {2023})}\BibitemShut {NoStop}%
\bibitem [{\citenamefont {Qian}\ \emph {et~al.}(2024)\citenamefont {Qian}, \citenamefont {Li}, \citenamefont {Li}, \citenamefont {Ren},\ and\ \citenamefont {Chen}}]{qian_deep_2024}%
  \BibitemOpen
  \bibfield  {author} {\bibinfo {author} {\bibfnamefont {Y.}~\bibnamefont {Qian}}, \bibinfo {author} {\bibfnamefont {X.}~\bibnamefont {Li}}, \bibinfo {author} {\bibfnamefont {Z.}~\bibnamefont {Li}}, \bibinfo {author} {\bibfnamefont {W.}~\bibnamefont {Ren}},\ and\ \bibinfo {author} {\bibfnamefont {J.}~\bibnamefont {Chen}},\ }\href {https://doi.org/10.48550/arXiv.2407.00707} {\bibinfo {title} {Deep learning quantum {{Monte Carlo}} for solids}} (\bibinfo {year} {2024}),\ \Eprint {https://arxiv.org/abs/2407.00707} {arXiv:2407.00707 [cond-mat, physics:physics]} \BibitemShut {NoStop}%
\bibitem [{\citenamefont {Carleo}\ and\ \citenamefont {Troyer}(2017)}]{carleo_solving_2017}%
  \BibitemOpen
  \bibfield  {author} {\bibinfo {author} {\bibfnamefont {G.}~\bibnamefont {Carleo}}\ and\ \bibinfo {author} {\bibfnamefont {M.}~\bibnamefont {Troyer}},\ }\href {https://doi.org/10.1126/science.aag2302} {\bibfield  {journal} {\bibinfo  {journal} {Science}\ }\textbf {\bibinfo {volume} {355}},\ \bibinfo {pages} {602} (\bibinfo {year} {2017})}\BibitemShut {NoStop}%
\bibitem [{\citenamefont {Vicentini}\ \emph {et~al.}(2019)\citenamefont {Vicentini}, \citenamefont {Biella}, \citenamefont {Regnault},\ and\ \citenamefont {Ciuti}}]{vicentini_variational_2019}%
  \BibitemOpen
  \bibfield  {author} {\bibinfo {author} {\bibfnamefont {F.}~\bibnamefont {Vicentini}}, \bibinfo {author} {\bibfnamefont {A.}~\bibnamefont {Biella}}, \bibinfo {author} {\bibfnamefont {N.}~\bibnamefont {Regnault}},\ and\ \bibinfo {author} {\bibfnamefont {C.}~\bibnamefont {Ciuti}},\ }\href {https://doi.org/10.1103/PhysRevLett.122.250503} {\bibfield  {journal} {\bibinfo  {journal} {Phys. Rev. Lett.}\ }\textbf {\bibinfo {volume} {122}},\ \bibinfo {pages} {250503} (\bibinfo {year} {2019})}\BibitemShut {NoStop}%
\bibitem [{\citenamefont {Han}\ \emph {et~al.}(2019)\citenamefont {Han}, \citenamefont {Zhang},\ and\ \citenamefont {E}}]{han_deepwf_2019}%
  \BibitemOpen
  \bibfield  {author} {\bibinfo {author} {\bibfnamefont {J.}~\bibnamefont {Han}}, \bibinfo {author} {\bibfnamefont {L.}~\bibnamefont {Zhang}},\ and\ \bibinfo {author} {\bibfnamefont {W.}~\bibnamefont {E}},\ }\href {https://doi.org/10.1016/j.jcp.2019.108929} {\bibfield  {journal} {\bibinfo  {journal} {Journal of Computational Physics}\ }\textbf {\bibinfo {volume} {399}},\ \bibinfo {pages} {108929} (\bibinfo {year} {2019})}\BibitemShut {NoStop}%
\bibitem [{\citenamefont {Pfau}\ \emph {et~al.}(2020)\citenamefont {Pfau}, \citenamefont {Spencer}, \citenamefont {Matthews},\ and\ \citenamefont {Foulkes}}]{pfau_ferminet_2020}%
  \BibitemOpen
  \bibfield  {author} {\bibinfo {author} {\bibfnamefont {D.}~\bibnamefont {Pfau}}, \bibinfo {author} {\bibfnamefont {J.~S.}\ \bibnamefont {Spencer}}, \bibinfo {author} {\bibfnamefont {A.~G. D.~G.}\ \bibnamefont {Matthews}},\ and\ \bibinfo {author} {\bibfnamefont {W.~M.~C.}\ \bibnamefont {Foulkes}},\ }\href {https://doi.org/10.1103/PhysRevResearch.2.033429} {\bibfield  {journal} {\bibinfo  {journal} {Physical Review Research}\ }\textbf {\bibinfo {volume} {2}},\ \bibinfo {pages} {033429} (\bibinfo {year} {2020})}\BibitemShut {NoStop}%
\bibitem [{\citenamefont {Hermann}\ \emph {et~al.}(2020)\citenamefont {Hermann}, \citenamefont {Sch{\"a}tzle},\ and\ \citenamefont {No{\'e}}}]{hermann_paulinet_2020}%
  \BibitemOpen
  \bibfield  {author} {\bibinfo {author} {\bibfnamefont {J.}~\bibnamefont {Hermann}}, \bibinfo {author} {\bibfnamefont {Z.}~\bibnamefont {Sch{\"a}tzle}},\ and\ \bibinfo {author} {\bibfnamefont {F.}~\bibnamefont {No{\'e}}},\ }\href {https://doi.org/10.1038/s41557-020-0544-y} {\bibfield  {journal} {\bibinfo  {journal} {Nature Chemistry}\ }\textbf {\bibinfo {volume} {12}},\ \bibinfo {pages} {891} (\bibinfo {year} {2020})}\BibitemShut {NoStop}%
\bibitem [{\citenamefont {Choo}\ \emph {et~al.}(2020)\citenamefont {Choo}, \citenamefont {Mezzacapo},\ and\ \citenamefont {Carleo}}]{choo_fermionic_2020}%
  \BibitemOpen
  \bibfield  {author} {\bibinfo {author} {\bibfnamefont {K.}~\bibnamefont {Choo}}, \bibinfo {author} {\bibfnamefont {A.}~\bibnamefont {Mezzacapo}},\ and\ \bibinfo {author} {\bibfnamefont {G.}~\bibnamefont {Carleo}},\ }\href {https://doi.org/10.1038/s41467-020-15724-9} {\bibfield  {journal} {\bibinfo  {journal} {Nature Communications}\ }\textbf {\bibinfo {volume} {11}},\ \bibinfo {pages} {2368} (\bibinfo {year} {2020})}\BibitemShut {NoStop}%
\bibitem [{\citenamefont {{von Glehn}}\ \emph {et~al.}(2023)\citenamefont {{von Glehn}}, \citenamefont {Spencer},\ and\ \citenamefont {Pfau}}]{glehn_psiformer_2023}%
  \BibitemOpen
  \bibfield  {author} {\bibinfo {author} {\bibfnamefont {I.}~\bibnamefont {{von Glehn}}}, \bibinfo {author} {\bibfnamefont {J.~S.}\ \bibnamefont {Spencer}},\ and\ \bibinfo {author} {\bibfnamefont {D.}~\bibnamefont {Pfau}},\ }in\ \href@noop {} {\emph {\bibinfo {booktitle} {The Eleventh International Conference on Learning Representations, {{ICLR}} 2023}}}\ (\bibinfo  {publisher} {OpenReview.net},\ \bibinfo {address} {kigali, rwanda},\ \bibinfo {year} {2023})\BibitemShut {NoStop}%
\bibitem [{\citenamefont {Yoshioka}\ \emph {et~al.}(2021)\citenamefont {Yoshioka}, \citenamefont {Mizukami},\ and\ \citenamefont {Nori}}]{yoshioka_rbm-solid_2021}%
  \BibitemOpen
  \bibfield  {author} {\bibinfo {author} {\bibfnamefont {N.}~\bibnamefont {Yoshioka}}, \bibinfo {author} {\bibfnamefont {W.}~\bibnamefont {Mizukami}},\ and\ \bibinfo {author} {\bibfnamefont {F.}~\bibnamefont {Nori}},\ }\href {https://doi.org/10.1038/s42005-021-00609-0} {\bibfield  {journal} {\bibinfo  {journal} {Communications Physics}\ }\textbf {\bibinfo {volume} {4}},\ \bibinfo {pages} {1} (\bibinfo {year} {2021})}\BibitemShut {NoStop}%
\bibitem [{\citenamefont {Li}\ \emph {et~al.}(2024{\natexlab{a}})\citenamefont {Li}, \citenamefont {Qian},\ and\ \citenamefont {Chen}}]{li_electric_2024}%
  \BibitemOpen
  \bibfield  {author} {\bibinfo {author} {\bibfnamefont {X.}~\bibnamefont {Li}}, \bibinfo {author} {\bibfnamefont {Y.}~\bibnamefont {Qian}},\ and\ \bibinfo {author} {\bibfnamefont {J.}~\bibnamefont {Chen}},\ }\href {https://doi.org/10.1103/PhysRevLett.132.176401} {\bibfield  {journal} {\bibinfo  {journal} {Physical Review Letters}\ }\textbf {\bibinfo {volume} {132}},\ \bibinfo {pages} {176401} (\bibinfo {year} {2024}{\natexlab{a}})}\BibitemShut {NoStop}%
\bibitem [{\citenamefont {Li}\ \emph {et~al.}(2022)\citenamefont {Li}, \citenamefont {Li},\ and\ \citenamefont {Chen}}]{li_deepsolid_2022}%
  \BibitemOpen
  \bibfield  {author} {\bibinfo {author} {\bibfnamefont {X.}~\bibnamefont {Li}}, \bibinfo {author} {\bibfnamefont {Z.}~\bibnamefont {Li}},\ and\ \bibinfo {author} {\bibfnamefont {J.}~\bibnamefont {Chen}},\ }\href {https://doi.org/10.1038/s41467-022-35627-1} {\bibfield  {journal} {\bibinfo  {journal} {Nature Communications}\ }\textbf {\bibinfo {volume} {13}},\ \bibinfo {pages} {7895} (\bibinfo {year} {2022})}\BibitemShut {NoStop}%
\bibitem [{\citenamefont {Wilson}\ \emph {et~al.}(2023)\citenamefont {Wilson}, \citenamefont {Moroni}, \citenamefont {Holzmann}, \citenamefont {Gao}, \citenamefont {Wudarski}, \citenamefont {Vegge},\ and\ \citenamefont {Bhowmik}}]{wilson_wapnet_2023}%
  \BibitemOpen
  \bibfield  {author} {\bibinfo {author} {\bibfnamefont {M.}~\bibnamefont {Wilson}}, \bibinfo {author} {\bibfnamefont {S.}~\bibnamefont {Moroni}}, \bibinfo {author} {\bibfnamefont {M.}~\bibnamefont {Holzmann}}, \bibinfo {author} {\bibfnamefont {N.}~\bibnamefont {Gao}}, \bibinfo {author} {\bibfnamefont {F.}~\bibnamefont {Wudarski}}, \bibinfo {author} {\bibfnamefont {T.}~\bibnamefont {Vegge}},\ and\ \bibinfo {author} {\bibfnamefont {A.}~\bibnamefont {Bhowmik}},\ }\href {https://doi.org/10.1103/PhysRevB.107.235139} {\bibfield  {journal} {\bibinfo  {journal} {Physical Review B}\ }\textbf {\bibinfo {volume} {107}},\ \bibinfo {pages} {235139} (\bibinfo {year} {2023})}\BibitemShut {NoStop}%
\bibitem [{\citenamefont {Cassella}\ \emph {et~al.}(2023)\citenamefont {Cassella}, \citenamefont {Sutterud}, \citenamefont {Azadi}, \citenamefont {Drummond}, \citenamefont {Pfau}, \citenamefont {Spencer},\ and\ \citenamefont {Foulkes}}]{cassella_discovering_2023}%
  \BibitemOpen
  \bibfield  {author} {\bibinfo {author} {\bibfnamefont {G.}~\bibnamefont {Cassella}}, \bibinfo {author} {\bibfnamefont {H.}~\bibnamefont {Sutterud}}, \bibinfo {author} {\bibfnamefont {S.}~\bibnamefont {Azadi}}, \bibinfo {author} {\bibfnamefont {N.~D.}\ \bibnamefont {Drummond}}, \bibinfo {author} {\bibfnamefont {D.}~\bibnamefont {Pfau}}, \bibinfo {author} {\bibfnamefont {J.~S.}\ \bibnamefont {Spencer}},\ and\ \bibinfo {author} {\bibfnamefont {W.~M.~C.}\ \bibnamefont {Foulkes}},\ }\href {https://doi.org/10.1103/PhysRevLett.130.036401} {\bibfield  {journal} {\bibinfo  {journal} {Physical Review Letters}\ }\textbf {\bibinfo {volume} {130}},\ \bibinfo {pages} {036401} (\bibinfo {year} {2023})}\BibitemShut {NoStop}%
\bibitem [{\citenamefont {Kim}\ \emph {et~al.}(2024)\citenamefont {Kim}, \citenamefont {Pescia}, \citenamefont {Fore}, \citenamefont {Nys}, \citenamefont {Carleo}, \citenamefont {Gandolfi}, \citenamefont {{Hjorth-Jensen}},\ and\ \citenamefont {Lovato}}]{kim_ucf_2024}%
  \BibitemOpen
  \bibfield  {author} {\bibinfo {author} {\bibfnamefont {J.}~\bibnamefont {Kim}}, \bibinfo {author} {\bibfnamefont {G.}~\bibnamefont {Pescia}}, \bibinfo {author} {\bibfnamefont {B.}~\bibnamefont {Fore}}, \bibinfo {author} {\bibfnamefont {J.}~\bibnamefont {Nys}}, \bibinfo {author} {\bibfnamefont {G.}~\bibnamefont {Carleo}}, \bibinfo {author} {\bibfnamefont {S.}~\bibnamefont {Gandolfi}}, \bibinfo {author} {\bibfnamefont {M.}~\bibnamefont {{Hjorth-Jensen}}},\ and\ \bibinfo {author} {\bibfnamefont {A.}~\bibnamefont {Lovato}},\ }\href {https://doi.org/10.1038/s42005-024-01613-w} {\bibfield  {journal} {\bibinfo  {journal} {Communications Physics}\ }\textbf {\bibinfo {volume} {7}},\ \bibinfo {pages} {148} (\bibinfo {year} {2024})}\BibitemShut {NoStop}%
\bibitem [{\citenamefont {Li}\ \emph {et~al.}(2024{\natexlab{b}})\citenamefont {Li}, \citenamefont {Qian}, \citenamefont {Ren}, \citenamefont {Xu},\ and\ \citenamefont {Chen}}]{li_emergent_2024}%
  \BibitemOpen
  \bibfield  {author} {\bibinfo {author} {\bibfnamefont {X.}~\bibnamefont {Li}}, \bibinfo {author} {\bibfnamefont {Y.}~\bibnamefont {Qian}}, \bibinfo {author} {\bibfnamefont {W.}~\bibnamefont {Ren}}, \bibinfo {author} {\bibfnamefont {Y.}~\bibnamefont {Xu}},\ and\ \bibinfo {author} {\bibfnamefont {J.}~\bibnamefont {Chen}},\ }\href {https://doi.org/10.48550/arXiv.2406.11134} {\bibinfo {title} {Emergent {{Wigner}} phases in moir{\'e} superlattice from deep learning}} (\bibinfo {year} {2024}{\natexlab{b}}),\ \Eprint {https://arxiv.org/abs/2406.11134} {arXiv:2406.11134 [cond-mat, physics:physics]} \BibitemShut {NoStop}%
\bibitem [{\citenamefont {Luo}\ \emph {et~al.}(2024)\citenamefont {Luo}, \citenamefont {Dai},\ and\ \citenamefont {Fu}}]{luo_simulating_2024}%
  \BibitemOpen
  \bibfield  {author} {\bibinfo {author} {\bibfnamefont {D.}~\bibnamefont {Luo}}, \bibinfo {author} {\bibfnamefont {D.~D.}\ \bibnamefont {Dai}},\ and\ \bibinfo {author} {\bibfnamefont {L.}~\bibnamefont {Fu}},\ }\href {https://doi.org/10.48550/arXiv.2406.17645} {\bibinfo {title} {Simulating moir{\'e} quantum matter with neural network}} (\bibinfo {year} {2024}),\ \Eprint {https://arxiv.org/abs/2406.17645} {arXiv:2406.17645 [cond-mat]} \BibitemShut {NoStop}%
\bibitem [{\citenamefont {Haldane}(1983)}]{haldane_fractional_1983}%
  \BibitemOpen
  \bibfield  {author} {\bibinfo {author} {\bibfnamefont {F.~D.~M.}\ \bibnamefont {Haldane}},\ }\href {https://doi.org/10.1103/PhysRevLett.51.605} {\bibfield  {journal} {\bibinfo  {journal} {Phys. Rev. Lett.}\ }\textbf {\bibinfo {volume} {51}},\ \bibinfo {pages} {605} (\bibinfo {year} {1983})}\BibitemShut {NoStop}%
\bibitem [{\citenamefont {Yoshioka}\ \emph {et~al.}(1983)\citenamefont {Yoshioka}, \citenamefont {Halperin},\ and\ \citenamefont {Lee}}]{yoshioka_ground_1983}%
  \BibitemOpen
  \bibfield  {author} {\bibinfo {author} {\bibfnamefont {D.}~\bibnamefont {Yoshioka}}, \bibinfo {author} {\bibfnamefont {B.~I.}\ \bibnamefont {Halperin}},\ and\ \bibinfo {author} {\bibfnamefont {P.~A.}\ \bibnamefont {Lee}},\ }\href {https://doi.org/10.1103/PhysRevLett.50.1219} {\bibfield  {journal} {\bibinfo  {journal} {Phys. Rev. Lett.}\ }\textbf {\bibinfo {volume} {50}},\ \bibinfo {pages} {1219} (\bibinfo {year} {1983})}\BibitemShut {NoStop}%
\bibitem [{\citenamefont {Wen}\ and\ \citenamefont {Zee}(1992)}]{wen_shift_1992}%
  \BibitemOpen
  \bibfield  {author} {\bibinfo {author} {\bibfnamefont {X.~G.}\ \bibnamefont {Wen}}\ and\ \bibinfo {author} {\bibfnamefont {A.}~\bibnamefont {Zee}},\ }\href {https://doi.org/10.1103/PhysRevLett.69.953} {\bibfield  {journal} {\bibinfo  {journal} {Physical Review Letters}\ }\textbf {\bibinfo {volume} {69}},\ \bibinfo {pages} {953} (\bibinfo {year} {1992})}\BibitemShut {NoStop}%
\bibitem [{\citenamefont {Jain}(2007)}]{jain_composite_2007}%
  \BibitemOpen
  \bibfield  {author} {\bibinfo {author} {\bibfnamefont {J.~K.}\ \bibnamefont {Jain}},\ }\href@noop {} {\emph {\bibinfo {title} {Composite {{Fermions}}}}},\ \bibinfo {edition} {1st}\ ed.\ (\bibinfo  {publisher} {Cambridge University Press},\ \bibinfo {address} {Cambridge ; New York},\ \bibinfo {year} {2007})\BibitemShut {NoStop}%
\bibitem [{\citenamefont {Morf}\ \emph {et~al.}(1986)\citenamefont {Morf}, \citenamefont {{d'Ambrumenil}},\ and\ \citenamefont {Halperin}}]{morf_microscopic_1986}%
  \BibitemOpen
  \bibfield  {author} {\bibinfo {author} {\bibfnamefont {R.}~\bibnamefont {Morf}}, \bibinfo {author} {\bibfnamefont {N.}~\bibnamefont {{d'Ambrumenil}}},\ and\ \bibinfo {author} {\bibfnamefont {B.~I.}\ \bibnamefont {Halperin}},\ }\href {https://doi.org/10.1103/PhysRevB.34.3037} {\bibfield  {journal} {\bibinfo  {journal} {Phys. Rev. B}\ }\textbf {\bibinfo {volume} {34}},\ \bibinfo {pages} {3037} (\bibinfo {year} {1986})}\BibitemShut {NoStop}%
\bibitem [{\citenamefont {Kato}(1957)}]{kato_cusp_1957}%
  \BibitemOpen
  \bibfield  {author} {\bibinfo {author} {\bibfnamefont {T.}~\bibnamefont {Kato}},\ }\href {https://doi.org/10.1002/cpa.3160100201} {\bibfield  {journal} {\bibinfo  {journal} {Communications on Pure and Applied Mathematics}\ }\textbf {\bibinfo {volume} {10}},\ \bibinfo {pages} {151} (\bibinfo {year} {1957})}\BibitemShut {NoStop}%
\bibitem [{\citenamefont {Wu}\ and\ \citenamefont {Yang}(1976)}]{wu_dirac_1976}%
  \BibitemOpen
  \bibfield  {author} {\bibinfo {author} {\bibfnamefont {T.~T.}\ \bibnamefont {Wu}}\ and\ \bibinfo {author} {\bibfnamefont {C.~N.}\ \bibnamefont {Yang}},\ }\href {https://doi.org/10.1016/0550-3213(76)90143-7} {\bibfield  {journal} {\bibinfo  {journal} {Nucl. Phys. B}\ }\textbf {\bibinfo {volume} {107}},\ \bibinfo {pages} {365} (\bibinfo {year} {1976})}\BibitemShut {NoStop}%
\bibitem [{\citenamefont {Martens}\ and\ \citenamefont {Grosse}(2015)}]{martens_kfac_2015}%
  \BibitemOpen
  \bibfield  {author} {\bibinfo {author} {\bibfnamefont {J.}~\bibnamefont {Martens}}\ and\ \bibinfo {author} {\bibfnamefont {R.}~\bibnamefont {Grosse}},\ }in\ \href@noop {} {\emph {\bibinfo {booktitle} {Proceedings of the 32nd {{International Conference}} on {{Machine Learning}}}}}\ (\bibinfo  {publisher} {PMLR},\ \bibinfo {year} {2015})\ pp.\ \bibinfo {pages} {2408--2417}\BibitemShut {NoStop}%
\bibitem [{\citenamefont {Jain}(1989)}]{jain_composite-fermion_1989}%
  \BibitemOpen
  \bibfield  {author} {\bibinfo {author} {\bibfnamefont {J.~K.}\ \bibnamefont {Jain}},\ }\href {https://doi.org/10.1103/PhysRevLett.63.199} {\bibfield  {journal} {\bibinfo  {journal} {Phys. Rev. Lett.}\ }\textbf {\bibinfo {volume} {63}},\ \bibinfo {pages} {199} (\bibinfo {year} {1989})}\BibitemShut {NoStop}%
\bibitem [{\citenamefont {Kamilla}\ \emph {et~al.}(1997)\citenamefont {Kamilla}, \citenamefont {Jain},\ and\ \citenamefont {Girvin}}]{kamilla_fermi-sea-like_1997}%
  \BibitemOpen
  \bibfield  {author} {\bibinfo {author} {\bibfnamefont {R.~K.}\ \bibnamefont {Kamilla}}, \bibinfo {author} {\bibfnamefont {J.~K.}\ \bibnamefont {Jain}},\ and\ \bibinfo {author} {\bibfnamefont {S.~M.}\ \bibnamefont {Girvin}},\ }\href {https://doi.org/10.1103/PhysRevB.56.12411} {\bibfield  {journal} {\bibinfo  {journal} {Physical Review B}\ }\textbf {\bibinfo {volume} {56}},\ \bibinfo {pages} {12411} (\bibinfo {year} {1997})}\BibitemShut {NoStop}%
\bibitem [{\citenamefont {Yannouleas}\ and\ \citenamefont {Landman}(2003)}]{yannouleas_two-dimensional_2003}%
  \BibitemOpen
  \bibfield  {author} {\bibinfo {author} {\bibfnamefont {C.}~\bibnamefont {Yannouleas}}\ and\ \bibinfo {author} {\bibfnamefont {U.}~\bibnamefont {Landman}},\ }\href {https://doi.org/10.1103/PhysRevB.68.035326} {\bibfield  {journal} {\bibinfo  {journal} {Phys. Rev. B}\ }\textbf {\bibinfo {volume} {68}},\ \bibinfo {pages} {035326} (\bibinfo {year} {2003})}\BibitemShut {NoStop}%
\bibitem [{\citenamefont {Yannouleas}\ and\ \citenamefont {Landman}(2002)}]{Yannouleas_trial_2002}%
  \BibitemOpen
  \bibfield  {author} {\bibinfo {author} {\bibfnamefont {C.}~\bibnamefont {Yannouleas}}\ and\ \bibinfo {author} {\bibfnamefont {U.}~\bibnamefont {Landman}},\ }\href {https://doi.org/10.1103/PhysRevB.66.115315} {\bibfield  {journal} {\bibinfo  {journal} {Phys. Rev. B}\ }\textbf {\bibinfo {volume} {66}},\ \bibinfo {pages} {115315} (\bibinfo {year} {2002})}\BibitemShut {NoStop}%
\bibitem [{\citenamefont {Shibata}\ and\ \citenamefont {Yoshioka}(2001)}]{shibata_ground-state_dmrg_2001}%
  \BibitemOpen
  \bibfield  {author} {\bibinfo {author} {\bibfnamefont {N.}~\bibnamefont {Shibata}}\ and\ \bibinfo {author} {\bibfnamefont {D.}~\bibnamefont {Yoshioka}},\ }\href {https://doi.org/10.1103/PhysRevLett.86.5755} {\bibfield  {journal} {\bibinfo  {journal} {Phys. Rev. Lett.}\ }\textbf {\bibinfo {volume} {86}},\ \bibinfo {pages} {5755} (\bibinfo {year} {2001})}\BibitemShut {NoStop}%
\bibitem [{\citenamefont {Thomson}(1904)}]{thomson_structure_1904}%
  \BibitemOpen
  \bibfield  {author} {\bibinfo {author} {\bibfnamefont {J.~J.}\ \bibnamefont {Thomson}},\ }\href@noop {} {\bibfield  {journal} {\bibinfo  {journal} {Phil. Mag.}\ }\textbf {\bibinfo {volume} {7}},\ \bibinfo {pages} {237} (\bibinfo {year} {1904})}\BibitemShut {NoStop}%
\bibitem [{\citenamefont {{E. Wigner}}(1934)}]{e_wigner_interaction_1934}%
  \BibitemOpen
  \bibfield  {author} {\bibinfo {author} {\bibnamefont {{E. Wigner}}},\ }\href@noop {} {\bibfield  {journal} {\bibinfo  {journal} {Phys. Rev.}\ }\textbf {\bibinfo {volume} {46}},\ \bibinfo {pages} {1002} (\bibinfo {year} {1934})}\BibitemShut {NoStop}%
\bibitem [{\citenamefont {Haldane}\ and\ \citenamefont {Rezayi}(1985)}]{haldane_finite-size_1985}%
  \BibitemOpen
  \bibfield  {author} {\bibinfo {author} {\bibfnamefont {F.~D.~M.}\ \bibnamefont {Haldane}}\ and\ \bibinfo {author} {\bibfnamefont {E.~H.}\ \bibnamefont {Rezayi}},\ }\href {https://doi.org/10.1103/PhysRevLett.54.237} {\bibfield  {journal} {\bibinfo  {journal} {Phys. Rev. Lett.}\ }\textbf {\bibinfo {volume} {54}},\ \bibinfo {pages} {237} (\bibinfo {year} {1985})}\BibitemShut {NoStop}%
\bibitem [{\citenamefont {Morf}\ \emph {et~al.}(2002)\citenamefont {Morf}, \citenamefont {{d'Ambrumenil}},\ and\ \citenamefont {Das~Sarma}}]{morf_excitation_2002}%
  \BibitemOpen
  \bibfield  {author} {\bibinfo {author} {\bibfnamefont {R.~H.}\ \bibnamefont {Morf}}, \bibinfo {author} {\bibfnamefont {N.}~\bibnamefont {{d'Ambrumenil}}},\ and\ \bibinfo {author} {\bibfnamefont {S.}~\bibnamefont {Das~Sarma}},\ }\href {https://doi.org/10.1103/PhysRevB.66.075408} {\bibfield  {journal} {\bibinfo  {journal} {Phys. Rev. B}\ }\textbf {\bibinfo {volume} {66}},\ \bibinfo {pages} {075408} (\bibinfo {year} {2002})}\BibitemShut {NoStop}%
\bibitem [{\citenamefont {Zhao}\ \emph {et~al.}(2022)\citenamefont {Zhao}, \citenamefont {Kudo}, \citenamefont {Faugno}, \citenamefont {Balram},\ and\ \citenamefont {Jain}}]{zhao_revisiting_2022}%
  \BibitemOpen
  \bibfield  {author} {\bibinfo {author} {\bibfnamefont {T.}~\bibnamefont {Zhao}}, \bibinfo {author} {\bibfnamefont {K.}~\bibnamefont {Kudo}}, \bibinfo {author} {\bibfnamefont {W.~N.}\ \bibnamefont {Faugno}}, \bibinfo {author} {\bibfnamefont {A.~C.}\ \bibnamefont {Balram}},\ and\ \bibinfo {author} {\bibfnamefont {J.~K.}\ \bibnamefont {Jain}},\ }\href {https://doi.org/10.1103/PhysRevB.105.205147} {\bibfield  {journal} {\bibinfo  {journal} {Phys. Rev. B}\ }\textbf {\bibinfo {volume} {105}},\ \bibinfo {pages} {205147} (\bibinfo {year} {2022})}\BibitemShut {NoStop}%
\bibitem [{\citenamefont {Boebinger}\ \emph {et~al.}(1985)\citenamefont {Boebinger}, \citenamefont {Chang}, \citenamefont {Stormer},\ and\ \citenamefont {Tsui}}]{boebinger_magnetic_1985}%
  \BibitemOpen
  \bibfield  {author} {\bibinfo {author} {\bibfnamefont {G.~S.}\ \bibnamefont {Boebinger}}, \bibinfo {author} {\bibfnamefont {A.~M.}\ \bibnamefont {Chang}}, \bibinfo {author} {\bibfnamefont {H.~L.}\ \bibnamefont {Stormer}},\ and\ \bibinfo {author} {\bibfnamefont {D.~C.}\ \bibnamefont {Tsui}},\ }\href {https://doi.org/10.1103/PhysRevLett.55.1606} {\bibfield  {journal} {\bibinfo  {journal} {Phys. Rev. Lett.}\ }\textbf {\bibinfo {volume} {55}},\ \bibinfo {pages} {1606} (\bibinfo {year} {1985})}\BibitemShut {NoStop}%
\bibitem [{\citenamefont {Du}\ \emph {et~al.}(1993)\citenamefont {Du}, \citenamefont {Stormer}, \citenamefont {Tsui}, \citenamefont {Pfeiffer},\ and\ \citenamefont {West}}]{du_experimental_1993}%
  \BibitemOpen
  \bibfield  {author} {\bibinfo {author} {\bibfnamefont {R.~R.}\ \bibnamefont {Du}}, \bibinfo {author} {\bibfnamefont {H.~L.}\ \bibnamefont {Stormer}}, \bibinfo {author} {\bibfnamefont {D.~C.}\ \bibnamefont {Tsui}}, \bibinfo {author} {\bibfnamefont {L.~N.}\ \bibnamefont {Pfeiffer}},\ and\ \bibinfo {author} {\bibfnamefont {K.~W.}\ \bibnamefont {West}},\ }\href {https://doi.org/10.1103/PhysRevLett.70.2944} {\bibfield  {journal} {\bibinfo  {journal} {Phys. Rev. Lett.}\ }\textbf {\bibinfo {volume} {70}},\ \bibinfo {pages} {2944} (\bibinfo {year} {1993})}\BibitemShut {NoStop}%
\bibitem [{\citenamefont {Halperin}\ \emph {et~al.}(1993)\citenamefont {Halperin}, \citenamefont {Lee},\ and\ \citenamefont {Read}}]{halperin_theory_1993}%
  \BibitemOpen
  \bibfield  {author} {\bibinfo {author} {\bibfnamefont {B.~I.}\ \bibnamefont {Halperin}}, \bibinfo {author} {\bibfnamefont {P.~A.}\ \bibnamefont {Lee}},\ and\ \bibinfo {author} {\bibfnamefont {N.}~\bibnamefont {Read}},\ }\href {https://doi.org/10.1103/PhysRevB.47.7312} {\bibfield  {journal} {\bibinfo  {journal} {Phys. Rev. B}\ }\textbf {\bibinfo {volume} {47}},\ \bibinfo {pages} {7312} (\bibinfo {year} {1993})}\BibitemShut {NoStop}%
\bibitem [{\citenamefont {Murthy}\ and\ \citenamefont {Shankar}(2003)}]{murthy_cs-review_2003}%
  \BibitemOpen
  \bibfield  {author} {\bibinfo {author} {\bibfnamefont {G.}~\bibnamefont {Murthy}}\ and\ \bibinfo {author} {\bibfnamefont {R.}~\bibnamefont {Shankar}},\ }\href {https://doi.org/10.1103/RevModPhys.75.1101} {\bibfield  {journal} {\bibinfo  {journal} {Reviews of Modern Physics}\ }\textbf {\bibinfo {volume} {75}},\ \bibinfo {pages} {1101} (\bibinfo {year} {2003})}\BibitemShut {NoStop}%
\bibitem [{\citenamefont {Yu}\ \emph {et~al.}(1998)\citenamefont {Yu}, \citenamefont {Su},\ and\ \citenamefont {Dai}}]{yu_effective_1998}%
  \BibitemOpen
  \bibfield  {author} {\bibinfo {author} {\bibfnamefont {Y.}~\bibnamefont {Yu}}, \bibinfo {author} {\bibfnamefont {Z.-B.}\ \bibnamefont {Su}},\ and\ \bibinfo {author} {\bibfnamefont {X.}~\bibnamefont {Dai}},\ }\href {https://doi.org/10.1103/PhysRevB.57.9897} {\bibfield  {journal} {\bibinfo  {journal} {Phys. Rev. B}\ }\textbf {\bibinfo {volume} {57}},\ \bibinfo {pages} {9897} (\bibinfo {year} {1998})}\BibitemShut {NoStop}%
\bibitem [{\citenamefont {Praz}(2007)}]{Praz_effective_2007}%
  \BibitemOpen
  \bibfield  {author} {\bibinfo {author} {\bibfnamefont {A.}~\bibnamefont {Praz}},\ }\href {https://doi.org/10.1103/PhysRevB.75.205342} {\bibfield  {journal} {\bibinfo  {journal} {Phys. Rev. B}\ }\textbf {\bibinfo {volume} {75}},\ \bibinfo {pages} {205342} (\bibinfo {year} {2007})}\BibitemShut {NoStop}%
\bibitem [{\citenamefont {{Melik-Alaverdian}}\ and\ \citenamefont {Bonesteel}(1995)}]{melik-alaverdian_composite_1995}%
  \BibitemOpen
  \bibfield  {author} {\bibinfo {author} {\bibfnamefont {V.}~\bibnamefont {{Melik-Alaverdian}}}\ and\ \bibinfo {author} {\bibfnamefont {N.~E.}\ \bibnamefont {Bonesteel}},\ }\href {https://doi.org/10.1103/PhysRevB.52.R17032} {\bibfield  {journal} {\bibinfo  {journal} {Phys. Rev. B}\ }\textbf {\bibinfo {volume} {52}},\ \bibinfo {pages} {R17032} (\bibinfo {year} {1995})}\BibitemShut {NoStop}%
\bibitem [{\citenamefont {Murthy}\ and\ \citenamefont {Shankar}(2002)}]{murthy_hamiltonian_2002}%
  \BibitemOpen
  \bibfield  {author} {\bibinfo {author} {\bibfnamefont {G.}~\bibnamefont {Murthy}}\ and\ \bibinfo {author} {\bibfnamefont {R.}~\bibnamefont {Shankar}},\ }\href {https://doi.org/10.1103/PhysRevB.65.245309} {\bibfield  {journal} {\bibinfo  {journal} {Physical Review B}\ }\textbf {\bibinfo {volume} {65}},\ \bibinfo {pages} {245309} (\bibinfo {year} {2002})}\BibitemShut {NoStop}%
\bibitem [{\citenamefont {Teng}\ \emph {et~al.}(2024)\citenamefont {Teng}, \citenamefont {Dai},\ and\ \citenamefont {Fu}}]{teng_solving_2024}%
  \BibitemOpen
  \bibfield  {author} {\bibinfo {author} {\bibfnamefont {Y.}~\bibnamefont {Teng}}, \bibinfo {author} {\bibfnamefont {D.~D.}\ \bibnamefont {Dai}},\ and\ \bibinfo {author} {\bibfnamefont {L.}~\bibnamefont {Fu}},\ }\href {https://doi.org/10.48550/arXiv.2412.00618} {\bibinfo {title} {Solving and visualizing fractional quantum {{Hall}} wavefunctions with neural network}} (\bibinfo {year} {2024}),\ \Eprint {https://arxiv.org/abs/2412.00618} {arXiv:2412.00618 [cond-mat]} \BibitemShut {NoStop}%
\bibitem [{dia()}]{diagham}%
  \BibitemOpen
  \href@noop {} {\bibinfo {title} {{{DiagHam}}}},\ \bibinfo {howpublished} {https://nick-ux.org/diagham}\BibitemShut {NoStop}%
\bibitem [{\citenamefont {Fishman}\ \emph {et~al.}(2022)\citenamefont {Fishman}, \citenamefont {White},\ and\ \citenamefont {Stoudenmire}}]{itensor}%
  \BibitemOpen
  \bibfield  {author} {\bibinfo {author} {\bibfnamefont {M.}~\bibnamefont {Fishman}}, \bibinfo {author} {\bibfnamefont {S.~R.}\ \bibnamefont {White}},\ and\ \bibinfo {author} {\bibfnamefont {E.~M.}\ \bibnamefont {Stoudenmire}},\ }\href {https://doi.org/10.21468/SciPostPhysCodeb.4} {\bibfield  {journal} {\bibinfo  {journal} {SciPost Phys. Codebases}\ ,\ \bibinfo {pages} {4}} (\bibinfo {year} {2022})}\BibitemShut {NoStop}%
\end{thebibliography}%

\clearpage

\appendix*

\section{Appendix: methodological and computational details}
\subsection{The role of monopole harmonics}

It is mentioned in the main text that LLL monopole harmonics $u_j^{Q+m} v_j^{Q-m}$ are used to avoid the divergence issue near the poles.
The monopole harmonics handle the complex phases arising from the Dirac string, while the neural network encodes the correlations between electrons.

To understand it, let us first consider the case without the Dirac string, where divergences in local energy near the poles also exist.
Consider a wavefunction $\psi \propto \theta^n \mathrm{e}^{\mathrm{i} m \phi}$ near the north pole.
To avoid divergence in local energy, the condition $m = \pm n$ must be satisfied.
Similarly, this condition must also hold near the south pole.
While a neural network wavefunction may not strictly satisfy this relation, it is generally not a significant problem in practice.
That is because the wavefunction should vanish rapidly near the poles as $|m|$ increases, contributing negligibly to the Monte Carlo sampling and the total energy.

With the Dirac string, the condition near the north pole becomes $m = Q \pm n$, and near the south pole, it becomes $m = -Q \pm n$.
A neural network wavefunction may not satisfy this relation, not to mention the singularities with ill-defined phases at the poles when $|m| = |Q|$.
Moreover, examining the monopole harmonics $u_j^{Q+m} v_j^{Q-m}$ reveals that orbitals with more complex phases (i.e., larger $|m|$) contribute more significantly near the poles.
This means the neural network must precisely capture the complex phase in these regions. 
Additionally, since the largest $|m|$ is proportional to $N$ for a fixed filling, the phase becomes even more complex and challenging to learn compared to the $Q=0$ case, where the largest $|m|$ is proportional to $N^2$.
By including LLL monopole harmonics in our neural network wavefunction, the complex phases of the wavefunction arising from the Dirac string can be properly handled.

\subsection{Neural network architecture}

In the neural network, $f_{k}(\mathbf{r}_j;\{\mathbf{r}_{\ne j}\})$, 
the input feature vector $\mathbf{f}_{i}^0$ for the electron $i$ is chosen as the Cartesian coordinates of the electron:
\begin{equation}
    \mathbf{f}_i^0 = [\sin \theta_i \cos \phi_i, \sin\theta_i\sin\phi_i, \cos\theta_i].
\end{equation}
The input features are then mapped to the attention inputs $\mathbf{h}_i^0$ by a linear projection $\mathbf{W}^0\mathbf{f}_i^0$.
After that, $\mathbf{h}_i^0$ is passed though a sequence of multi-head attention layers and fully connected layers with residual connection:
\begin{multline}
    \mathbf{f}_i^{l+1} = \mathbf{h}_i^l + \mathbf{W}_o^l \operatorname{concat}_h [\\
    \text{\sc{SelfAttn}}_i (\mathbf{h}_1^l,\dots,\mathbf{h}_N^l;\mathbf{W}_q^{lh},\mathbf{W}_k^{lh},\mathbf{W}_v^{lh})],
\end{multline}
\begin{equation}
    \mathbf{h}_i^{l+1} = \mathbf{f}_i^{l+1} + \tanh (\mathbf{W}^{l+1}\mathbf{f}_i^{l+1} + \mathbf{b}^{l+1}),
\end{equation}
where $l$ indexes neural network layers and $h$ indexes heads of the attention layer.
And the standard self-attention is defined as:
\begin{multline}
    \text{\sc{SelfAttn}}_i (\mathbf{h}_1,\dots,\mathbf{h}_N;\mathbf{W}_q,\mathbf{W}_k,\mathbf{W}_v)=\\
    \frac{1}{\sqrt{d}}\sum_j \sigma_j \left( \mathbf{q}_1^\mathsf{T} \mathbf{k}_i,\dots,\mathbf{q}_N^\mathsf{T} \mathbf{k}_i \right) \mathbf{v}_j,
\end{multline}
\begin{gather}
    \mathbf{k}_i = \mathbf{W}_k \mathbf{h}_i,\quad
    \mathbf{q}_i = \mathbf{W}_q \mathbf{h}_i,\quad 
    \mathbf{v}_i = \mathbf{W}_v \mathbf{h}_i,\\
    \sigma_i(x_1,\dots,x_N)=\frac{\exp(x_i)}{\sum_j \exp(x_j)},
\end{gather}
where $d$ is the output dimension of the key and query weights.
The hyperparameters of the neural network used in this study are listed in Table S1.

\subsection{Energy corrections}

The energy results shown in the main text include the contribution from background charge,
\begin{multline}\label{eq:bg}
    E_\text{bg}=E_\text{el--bg}+E_\text{bg--bg}\\
    = \kappa \hbar \omega_c \ell \left(-\frac{N^2}{R} + \frac{N^2}{2R}\right)
    = -\kappa \hbar \omega_c \frac{N^2}{2\sqrt{Q}}.
\end{multline}
In addition, we shift the energy by $N\omega_c/2$ and apply a density correction to reduce the dependence of the energy on the system size.
The corrected energy $E_c$ is then defined as:
\begin{equation}\label{eq:density-correction}
    E_c = \sqrt{\frac{2Q\nu}{N}}\left(E_v + E_\text{bg}-\frac{N\omega_c}{2}\right).
\end{equation}

Notably, the background contribution to the excited states differs from that of the ground state.
Specifically, for the $\nu=1/3$ state, these contributions are given by:
\begin{equation}
    E_\text{bg}^\text{qp/qh}=-\kappa \hbar \omega_c \frac{N^2-q^2}{2\sqrt{Q^\text{qp/qh}}},
\end{equation}
where $|q|=1/3$ corresponds to the excess charge in the excited states.
And the transport gap after correction $E^\text{gap}_c$ is essentially
\begin{multline}
    E^\text{gap}_c = \sqrt{\frac{2Q^\text{qp}\nu}{N}}\left(E_v^\text{qp} + E_\text{bg}^\text{qp} - \frac{N\omega_c}{2}\right)\\
    + \sqrt{\frac{2Q^\text{qh}\nu}{N}}\left(E_v^\text{qh} + E_\text{bg}^\text{qh} - \frac{N\omega_c}{2}\right)
    - 2 E_c.
\end{multline}

\subsection{Computational details}

\paragraph{Overlap with Laughlin wavefunction ---}
The Laughlin wavefunction for $\nu=1/m$ state on a sphere is given by:
\begin{equation}
\psi^{\text{Laughlin}} = \prod_{i<j} (u_i v_j - u_j v_i)^m,
\end{equation}
where $u_i=\cos\frac{\theta_i}{2} \mathrm{e}^{\mathrm{i} \phi_i/2}$ and $v_i=\sin\frac{\theta_i}{2} \mathrm{e}^{-\mathrm{i} \phi_i/2}$ are the spinor coordinates of the $i$-th electron, and $m$ is an odd integer.
The overlap between the neural network wavefunction $\psi^\text{nn}$ and $\psi^\text{Laughlin}$ is defined in Eq.~\eqref{eq:overlap}.
Given the two wavefunctions are similar,  the overlap can be efficiently sampled using importance sampling from the distribution $|\psi^\text{nn}|^2$, expressed as:
\begin{equation}
    |S|^2 = \frac{\left|\left\langle \psi^\text{Laughlin}/\psi^\text{nn} \right\rangle\right|^2}{\left\langle\left| \psi^\text{Laughlin}/\psi^\text{nn} \right|^2\right\rangle},
\end{equation}
where $\langle\cdot\rangle$ denote the expectation value under the distribution $|\psi^\text{nn}|^2$.

\paragraph{The number of electrons on the lowest Landau level}
can be calculated using the trace of the one-body reduced density matrix (1-RDM), with the basis consisting of LLL monopole harmonics $\varphi_i$. The 1-RDM for wavefunction $\psi$ is defined as:
\begin{multline}\label{eq:1rdm}
    \Gamma_{ij} = \sum_{a=1}^N \int \mathrm{d}^2\mathbf{r}_1 \cdots \mathrm{d}^2\mathbf{r}_N \mathrm{d}^2\mathbf{r}^\prime \Big[ \psi^*(\mathbf{r}_1, \dots ,\mathbf{r}_N) \varphi_i (\mathbf{r}_a)\\
    \varphi_j^* (\mathbf{r}^\prime) \psi(\mathbf{r}_1, \dots, \mathbf{r}_{a-1},\mathbf{r}^\prime,\mathbf{r}_{a+1}, \dots, \mathbf{r}_N) \Big].
\end{multline}
In practice, importance sampling based on the distribution $|\psi|^2$ is applied for the integral over $\mathrm{d}^2\mathbf{r}_1 \cdots \mathrm{d}^2\mathbf{r}_N$, and uniform sampling is used for $\mathrm{d}^2\mathbf{r}^\prime$. Assuming the monopole harmonics are normalized, the integral in Eq.~\eqref{eq:1rdm} can be calculated with:
\begin{equation}
    \left\langle \frac{\psi(\mathbf{r}_1, \dots, \mathbf{r}_{a-1},\mathbf{r}^\prime,\mathbf{r}_{a+1}, \dots, \mathbf{r}_N)}{\psi(\mathbf{r}_1, \dots ,\mathbf{r}_N)} \varphi_i (\mathbf{r}_a) \varphi_j^* (\mathbf{r}^\prime) \right\rangle.
\end{equation}

\paragraph{The pair correlation function} is defined as:
\begin{equation}
    g(\mathbf{r}) = \frac{1}{\rho N} \left\langle \sum_{i \ne j} \delta^{(2)} (\mathbf{r} - \mathbf{r}_i + \mathbf{r}_j) \right\rangle,
\end{equation}
where $\rho$ is the electron density on the sphere. Since $g(\mathbf{r}) = g(|\mathbf{r}|)$ on the spherical geometry, the PCF depends only on the angular separation $\theta_{ij}$ between electrons $i$ and $j$. Therefore, we can express the PCF as:
\begin{equation}
    g(\theta) = \frac{1}{\rho N} \left\langle \sum_{i \ne j} \delta (\theta - \theta_{ij}) \right\rangle.
\end{equation}
We approximate the $\delta$ function as:
\begin{equation}
    \delta(\theta) \approx \begin{cases}
    \dfrac{1}{\Delta S} & \text{if } |\theta| < \dfrac{\pi}{2K}, \\
    0 & \text{otherwise},
    \end{cases}
\end{equation}
where $K$ is the number of bins, and $\Delta S = 2\pi R^{2} \sin\theta \Delta\theta$ with $\Delta\theta = \pi/K$. Therefore, in Monte Carlo simulations, we collect $\theta_{ij}$ in bins with weight $1/\sin\theta_{ij}$, and multiply by an overall factor:
\begin{equation}
    \frac{1}{\rho N \Delta S} = \frac{4\pi R^{2}}{N^{2}} \frac{K}{2\pi^{2} R^{2} \sin\theta} = \frac{2K}{\pi N^{2}}.
\end{equation}

\paragraph{ED calculations} of energy are performed using the DiagHam library~\cite{diagham}.
The density profiles of quasiparticle and quasihole excitations are obtained by DMRG calculations in the LLL using the ITensor library for technical convenience~\cite{itensor}. We have verified that the energy values calculated by DMRG are identical to those obtained from DiagHam and the entanglement entropy is also converged to machine precision. These facts guarantee that the results from DMRG calculations are equivalent to ED calculations.

\end{document}


\title{Supplemental Materials: Taming Landau level mixing in fractional quantum Hall states with deep learning}

\author{Yubing Qian}
\affiliation{School of Physics, Peking University, Beijing 100871, People’s Republic of China}%
\affiliation{ByteDance Research China, Fangheng Fashion Center, No. 27, North 3rd Ring West Road, Haidian District, Beijing 100098, People’s Republic of China}

\author{Tongzhou Zhao}
\email{tzzhao\_2022@iphy.ac.cn}
\thanks{Y.Q. and T.Z. contributed equally to this work.}
\affiliation{Institute of Physics, Chinese Academy of Sciences, Beijing 100190, China}

\author{Jianxiao Zhang}
\affiliation{Department of Physics, 104 Davey Lab, Pennsylvania State University, University Park, Pennsylvania 16802, USA}

\author{Tao Xiang}
\affiliation{Institute of Physics, Chinese Academy of Sciences, Beijing 100190, China}

\author{Xiang Li}
\email{lixiang.62770689@bytedance.com}
\affiliation{ByteDance Research China, Fangheng Fashion Center, No. 27, North 3rd Ring West Road, Haidian District, Beijing 100098, People’s Republic of China}

\author{Ji Chen}
\email{ji.chen@pku.edu.cn}
\affiliation{School of Physics, Peking University, Beijing 100871, People’s Republic of China}
\affiliation{Interdisciplinary Institute of Light-Element Quantum Materials, Frontiers Science Center for Nano-Optoelectronics, Peking University, Beijing 100871, People’s Republic of China}

\date{\today}

\maketitle

\begin{table}
\caption{\label{tab:params}Neural network hyperparameters.}
\begin{tabular}{@{\quad} l @{\qquad} l @{\qquad} l @{\quad}}
\toprule
Source & Parameter & Value \\
\hline
\multirow{5}{*}{Network Architecture}
& Determinants & 1 \\
& Network layers & 4 \\
& Attention heads & 4 \\
& Attention dimension & 64 \\
& Fully connected layer dimension & 256 \\
\hline
Training in Fig. 2 & Iterations  & $1\times 10^5$ \\
Training in Fig. 3--4 & Iterations & $3\times 10^4$ \\
\multirow{2}{*}{Training in Fig. 4a}
& Additional iterations with penalty & $2\times10^4$ \\
& Penalty strength $\beta$ & $0.02$ \\
\multirow{2}{*}{Training in Fig. 4b}
& Additional iterations with penalty & $4\times10^4$ \\
& Penalty strength $\beta$ & $0.01$ \\
\botrule
\end{tabular}
\end{table}

\begin{table}
\caption{\label{tab:energy}Neural network and ED energy results with different number of electrons for $\nu = 1/3$ filling.
The energy ($E_c$) includes the contribution from background charge.
In addition, it is shifted by $N\omega_c/2$ and a density correction to reduce the system size dependence is applied.
The energy unit is $\hbar \omega_c \kappa$.
The ED result for 12 electrons is not available due to huge memory requirements.}
\begin{tabular}{@{\quad} l @{\qquad} l @{\qquad} l @{\quad}}
\toprule
$N$ & $E_c/N\kappa\hbar\omega_c$ & ED \\
\hline
6 & $-0.411688(3)$ & $-0.410950$ \\
7 & $-0.411547(3)$ & $-0.410819$ \\
8 & $-0.411444(3)$ & $-0.410736$ \\
9 & $-0.411369(3)$ & $-0.410679$ \\
10 & $-0.411280(3)$ & $-0.410629$ \\
11 & $-0.411202(4)$ & $-0.410583$ \\
12 & $-0.411124(4)$ & N/A \\
\botrule
\end{tabular}
\end{table}

\begin{table}
\caption{\label{tab:llm}Neural network and ED results for $\nu = 1/3$ filling with different $\kappa$,
including overlaps with Laughlin wavefunction, electrons on the lowest Landau level, and quasiparticle/quasihole excitation energies.
The overlap between ED and Laughlin wavefunction is not available.}
\begin{ruledtabular}
\begin{tabular}{lllllll}
$\kappa$ & Overlap & $N_\text{LLL}$ & $E_c/N\kappa\hbar\omega_c$ & $E_c^\text{qp}/N\kappa\hbar\omega_c$ & $E_c^\text{qh}/N\kappa\hbar\omega_c$ & $E_c^\text{gap}/N\kappa\hbar\omega_c$ \\
\hline
$0.5$ & $0.9976(2)$ & $5.9989(8)$ & $-0.41106(2)$ & $-0.397425(8)$ & $-0.405843(7)$ & $0.1130(3)$ \\
$1$ & $0.994(2)$ & $5.9955(8)$ & $-0.411645(3)$ & $-0.39897(1)$ & $-0.406667(3)$ & $0.1058(2)$ \\
$3$ & $0.975(2)$ & $5.9644(8)$ & $-0.413331(3)$ & $-0.403202(1)$ & $-0.408784(1)$ & $0.08805(5)$ \\
$10$ & $0.873(1)$ & $5.7500(8)$ & $-0.417425(1)$ & $-0.4105431(4)$ & $-0.4136791(4)$ & $0.06377(1)$ \\
\hline
ED & N/A & 6 & $-0.410950$ & $-0.396703$ & $-0.405834$ & $0.116170$ \\
\end{tabular}
\end{ruledtabular}
\end{table}

\begin{figure*}
    \centering
    \includegraphics{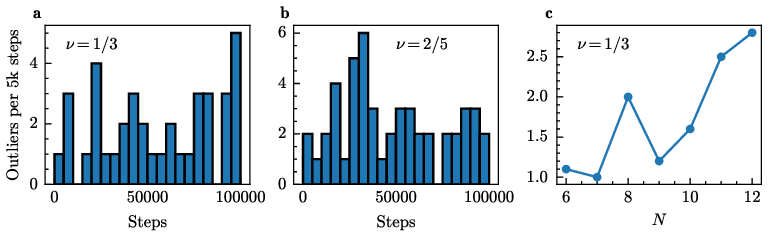}
    \caption{%
Although multiplying monopole harmonics with the neural network output eliminates most divergence problems, some outliers still occur during training.
Outliers are removed when plotting the training curves to provide a clearer view of the training process.
The removed outliers are rare and do not influence the overall training process.
\textbf{(a)} Number of outliers per 5000 steps when training $N=8$ electrons, $\nu=1/3$ filling, Landau level mixing parameter $\kappa=1$.
\textbf{(b)} Similar to (a), but for $\nu=2/5$ filling.
\textbf{(c)} Average number of outliers per 5000 training steps as a function of electron numbers for $\nu=1/3$ filling.
    }
    \label{fig:outliers}
\end{figure*}

\begin{figure*}
    \centering
    \includegraphics{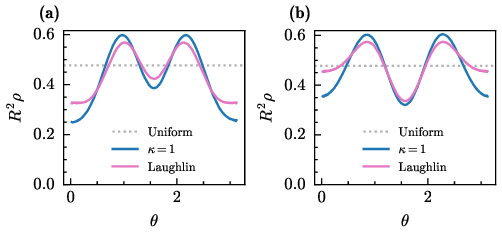}
    \caption{%
The charge densities showing the quasiparticle and quasihole excitations localized at the equator
\textbf{(a)} Density of quasiparticle states derived from the Laughlin and neural network wavefunctions.
The quasiparticle is localized at the equator.
The density is scaled by the square of the radius $R$ of the spherical geometry.
\textbf{(b)} Similar to (a), but for the density of quasihole states.
The densities obtained from the neural network wavefunction and the Laughlin wavefunction agree well, although the neural network results exhibit more fluctuations and are less localized.
This observation is consistent with the physics discussed in the main text.
Furthermore, due to the limited size of the sphere, the density shows less structure compared to excitations near the north pole.
    }
    \label{fig:equator-density}
\end{figure*}